\documentclass[fleqn,usenatbib]{mnras}

\usepackage{newtxtext,newtxmath}
\usepackage{comment}
\usepackage{graphicx}
\usepackage{float} 

\usepackage{longtable, booktabs}
\usepackage{threeparttable}
\usepackage{threeparttablex}
\usepackage{subcaption}
\usepackage{endnotes}

\usepackage[T1]{fontenc}

\DeclareRobustCommand{\VAN}[3]{#2}
\let\VANthebibliography\thebibliography
\def\thebibliography{\DeclareRobustCommand{\VAN}[3]{##3}\VANthebibliography}

\usepackage{array}
\usepackage{rotating}

\title[SN~2020jfo]{Photometric and spectroscopic analysis of the Type II SN~2020jfo with a short plateau}

\author[B. Ailawadhi et al.]{B. Ailawadhi,$^{1,2}$\thanks{E-mail: bhavya@aries.res.in}
R. Dastidar,$^{3,4}$
K. Misra,$^{1}$
R. Roy,$^{5,6}$
D. Hiramatsu,$^{7,8}$
D. A. Howell,$^{7,8}$
T. G. Brink,$^{9}$
W. Zheng,$^{9}$
\newauthor
L. Galbany,$^{10,11}$
M. Shahbandeh,$^{12}$
I. Arcavi,$^{13,14}$
C. Ashall,$^{15}$
K. A. Bostroem,$^{16}$
J. Burke,$^{7,8}$
T. Chapman,$^{9}$ 
\newauthor
Dimple,$^{1,2}$
A. V. Filippenko,$^{9,17}$
A. Gangopadhyay,$^{18}$
A. Ghosh,$^{1,19}$
A. M. Hoffman,$^{20,9,21}$
G. Hosseinzadeh,$^{22}$
\newauthor
C. Jennings,$^{9}$ 
V. K. Jha,$^{1,2}$
A. Kumar,$^{1,19,23}$
E. Karamehmetoglu,$^{24}$
C. McCully,$^{7,8}$
E. McGinness,$^{9}$ 
\newauthor
T. E. M\"uller-Bravo,$^{10}$
Y. S. Murakami,$^{21,9,25}$ 
S. B. Pandey,$^{1}$
C. Pellegrino,$^{7,8}$
L. Piscarreta,$^{10}$
J. Rho,$^{26}$
\newauthor
M. Stritzinger,$^{22}$
J. Sunseri,$^{9,21}$ 
S. D.~Van Dyk,$^{27}$
L. Yadav$^{2}$
\\
$^{1}$ Aryabhatta Research Institute of Observational Sciences, Manora Peak, Nainital 263 001, India\\
$^{2}$ Department of Physics, Deen Dayal Upadhyaya Gorakhpur University, Gorakhpur-273009, India\\
$^{3}$ Millennium Institute of Astrophysics, Nuncio Monsenor Sótero Sanz 100, Providencia, Santiago, 8320000 Chile\\
$^{4}$ Instituto de Astrofisica, Universidad Andres Bello, Fernandez Concha 700, Las Condes, Santiago, Chile\\
$^{5}$ Manipal Centre for Natural Sciences, Centre of Excellence, Manipal Academy of Higher Education, Manipal 576104, Karnataka, India\\
$^{6}$ Inter-University Center for Astronomy and Astrophysics, Ganeshkhind, Pune, Maharashtra-411007, India\\
$^{7}$ Department of Physics, University of California, Santa Barbara, CA 93106-9530, USA\\
$^{8}$ Las Cumbres Observatory, 6740 Cortona Drive, Suite 102, Goleta, CA 93117-5575, USA\\
$^{9}$ Department of Astronomy, University of California, Berkeley, CA 94720-3411, USA\\
$^{10}$ Institute of Space Sciences (ICE, CSIC), Campus UAB, Carrer de Can
Magrans, s/n, E-08193 Barcelona, Spain\\
$^{11}$ Institut d’Estudis Espacials de Catalunya (IEEC), E-08034 Barcelona,
Spain\\
$^{12}$ Department of Physics, Florida State University, 77 Chieftan Way, Tallahassee, FL 32306, USA\\
$^{13}$ The School of Physics and Astronomy, Tel Aviv University, Tel Aviv 69978, Israel\\
$^{14}$ CIFAR Azrieli Global Scholars program, CIFAR, Toronto, Canada\\
$^{15}$ Department of Physics, Florida State University, Tallahassee, FL 32306, USA\\
$^{16}$ DiRAC Institute, Department of Astronomy, University of Washington, Box 351580, U.W., Seattle, WA 98195, USA\\
$^{17}$ Miller Institute for Basic Research in Science, University of California, Berkeley, CA 94720, USA\\
$^{18}$ Hiroshima Astrophysical Science Centre, Hiroshima University, 1-3-1 Kagamiyama, Higashi-Hiroshima, Hiroshima 739-8526, Japan\\
$^{19}$ School of Studies in Physics and Astrophysics, Pandit Ravishankar Shukla University, Chattisgarh 492010, India\\
$^{20}$ Institute for Astronomy, University of Hawai‘i at Manoa, 2680 Woodlawn Dr., Honolulu, HI 96822, USA\\
$^{21}$ Department of Physics, University of California, Berkeley, CA 94720-7300, USA\\
$^{22}$ Steward Observatory, University of Arizona, 933 North Cherry Avenue, Tucson, AZ 85721-0065, USA\\
$^{23}$ Department of Physics, University of Warwick, Gibbet Hill Road, Coventry CV4 7AL, UK\\
$^{24}$ Department of Physics and Astronomy, Aarhus University, Ny Munkegade 120, DK-8000 Aarhus C, Denmark\\
$^{25}$ Department of Physics and Astronomy, Johns Hopkins University, Baltimore, MD 21218, USA\\
$^{26}$ SETI Institute, 189 North Bernardo Avenue, Suite 200, Mountain View, CA 94043, USA\\
$^{27}$ Caltech/IPAC, Mailcode 100-22, Pasadena, CA 91125, USA
}

\date{Accepted XXX. Received YYY; in original form ZZZ}

\pubyear{2021}

\begin{document}
\label{firstpage}
\pagerange{\pageref{firstpage}--\pageref{lastpage}}
\maketitle

\begin{abstract}
We present high-cadence photometric and spectroscopic observations of SN~2020jfo in ultraviolet and optical/near-infrared bands starting from $\sim 3$ to $\sim 434$ days after the explosion, including the earliest data with the 10.4\,m GTC. SN~2020jfo is a hydrogen-rich Type II SN with a relatively short plateau duration ($67.0 \pm 0.6$ days). When compared to other Type II supernovae (SNe) of similar or shorter plateau lengths, SN~2020jfo exhibits a fainter peak absolute $V$-band magnitude ($M_V = -16.90 \pm 0.34$ mag). SN~2020jfo shows significant H$\alpha$ absorption in the plateau phase similar to that of typical SNe~II. The emission line of stable [Ni~II] $\lambda$7378, mostly seen in low-luminosity SNe~II, is very prominent in the nebular-phase spectra of SN~2020jfo. Using the relative strengths of [Ni~II] $\lambda$7378 and [Fe~II] $\lambda$7155, we derive the Ni/Fe production (abundance) ratio of 0.08--0.10, which is $\sim 1.5$ times the solar value. The progenitor mass of SN~2020jfo from nebular-phase spectral modelling and semi-analytical modelling falls in the range of 12--15\,$M_\odot$. Furthermore, semi-analytical modelling suggests a massive H envelope in the progenitor of SN~2020jfo, which is unlikely for SNe~II having short plateaus.
\end{abstract}



\section{Introduction}
\label{introduction}

Core-collapse supernovae (CCSNe) represent the end stages of stars having zero-age main-sequence (ZAMS) masses $\gtrsim 8\,M_\odot$. These SNe emerge from the gravitational collapse of the degenerate core of massive stars. The presence of hydrogen (H) lines in the SN spectra around maximum brightness distinguishes Type II SNe from other subtypes of CCSNe \citep{filippenko}. The plateau (P) feature (which lasts for $\sim 80$--120 days after maximum brightness; \citealt{barbon}) and the linear (L) decline feature in the light curve characterises the events as Type IIP and Type IIL SNe, respectively.  However, as more SNe were discovered through systematic surveys, it became evident that typical SN II light-curve morphologies formed a continuous distribution \citep{anderson14}. Another subclass of SNe~II are the Type IIb, which initially show features of hydrogen and later transition to helium, suggesting small amounts of H in the ejecta \citep{filippenko1988}. Hereafter, we will refer to the hydrogen-rich class as Type II SNe. The H recombination phase will be referred to as the ``plateau." The Type II SNe with a plateau decline rate greater than 1 mag (100 day)$^{-1}$ will be referred to as the fast-declining SNe~II  and those with a decline rate less than 1 mag (100 day)$^{-1}$ will be referred to as the slow-declining SNe~II \citep{Faran2014}.

The formation of the plateau in SNe~II occurs due to energy deposition in the hydrogen-rich envelope of the star by the shock wave, which is gradually radiated away as the recombination wave moves inward in mass coordinates. The transition from plateau to the exponential decay/tail phase marks the end of the recombination phase. There is a drop of 1.0--2.6 mag from the plateau to the tail phase in the case of normal-luminosity SNe~II \citep{valenti16}. On the other hand, a 3--5 mag drop has been observed in the low-luminosity Type II SNe (e.g. $\sim 3.83$ mag in SN~2005cs; \citealt{tsvetkov,Spiro2013}). However, recent studies have reported the presence of outliers --- for example, a large drop of 3.8 mag was reported in SN~2018zd \citep{hiramatsu2018zd}, a normal-luminosity SN~II, and a drop of $\sim 2$ mag was observed in the low-luminosity Type II SN~2016afq \citep{Muller-Bravo2020}. After the drop, the SN enters the exponentially declining tail phase, which is powered by the decay of $^{56}$Co to $^{56}$Fe.

In pre-explosion imaging studies of the galaxies hosting these SNe, red supergiant (RSG) stars have been shown to be the progenitors of the slowly declining SNe~II, which preserve the hydrogen envelope at the time of explosion. The mass range of the slowly declining Type II SN progenitors estimated from pre-explosion imaging is 8--17\,\(M_\odot\), while the apparent lack of higher-mass progenitors (17--30\,\(M_\odot\)) gives rise to the ``RSG problem" \citep{smartt2009,smartt2015}.

In the recent study by \citet{anderson14} and \citet{valenti16}, the estimated parameters in a sample of SNe~II display a wide range of peak absolute magnitudes ($-14 > M_{V,{\rm peak}} > -19.5$ mag) and plateau luminosities ($-14 > M_{V,{\rm 50\,d}} > -18$ mag). The $^{56}$Ni mass is estimated to lie in the range 0.005--0.280\,\(M_\odot\) \citep{Muller2017,rodriguez} and the ejecta velocities lie between $9600 > v_{\rm ph} > 1500$  km s$^{-1}$ (\citealt{gutirrez}). This shows the remarkable diversity in the features of Type II SN population, encompassing faint, low-velocity, nickel-poor events such as SN~1997D \citep{benetti2001,chugai2000} to bright, high-velocity, nickel-rich events such as SN~1992am \citep{nadyozhin2003}. Moreover, some SNe~II are found to have a shorter plateau than normal (e.g., SNe 1995ad, 2006Y, 2006ai, 2007od, and 2016egz; \citealt{inserra2011, Inserra1995ad, hiramatsu}), while for some other SNe~II, longer plateaus have been reported (e.g., SNe~2009ib; \citealt{takts}, 2015ba; \citealt{dastidar}). The plateau length largely depends on the recombination time of all the ionized hydrogen, which is proportional to the mass of the hydrogen envelope \citep{popov93,kasen}. The analytical and numerical modelling of the light curves of Type II SNe suggests that the length of the plateau scales with the explosion and progenitor properties of SNe such as the initial stellar mass, metallicity, and explosion energy \citep{popov93,sukhbold,goldberg}.

\citet{hiramatsu} studied the properties of three SNe~II (SNe~2006Y, 2006ai, 2016egz) with shorter plateau lengths ranging from 50 to 70 days. They reported some properties of these three SNe which make them stand out from  normal Type IIP SNe. In their work, the three Type II SNe were identified as a transitional class between Type IIL and IIb SNe with a narrow H-rich envelope mass window ($M_{\rm H-env} \approx 1.7\,M_\odot$). The plateau phase of these three SNe~II has been claimed to be powered by $^{56}$Ni decay, unlike other Type II SNe. The photospheric luminosity and the effective temperature of these three SNe were found to lie in the range $5.23 < \log_{10}(L_\mathrm{ph}/L_\odot) < 5.53$ and $3.65 < \log_{10}(T_\mathrm{eff}/\mathrm{K}) < 3.72$ (respectively) from modelling, and these values correspond to luminous RSGs.

In this paper, we present the photometric and spectroscopic analysis of the Type II SN~2020jfo. The light curve and spectra of SN~2020jfo show similarity with normal SNe~II, but with a comparatively shorter plateau length in comparison to the normal events. The paper is structured as follows. The discovery of SN~2020jfo is presented in Section \ref{discovery}. Section \ref{observations} provides the details of photometric and spectroscopic observations and data-reduction techniques. The light curve and spectroscopic evolution are presented in Sections \ref{lightcurves} and \ref{spectroscopy}, respectively. The properties of the progenitor and explosion are discussed in Section \ref{progenitor}. A comparison of SN~2020jfo with other Type II SNe is given in Section \ref{collocation}, and our conclusions are presented in Section \ref{conclusions}.

\section{SN~2020jfo Discovery}
\label{discovery}

SN~2020jfo (also known as ZTF20aaynrrh) was discovered on 2020-05-06 (UTC dates are used throughout this paper; 58975.2 MJD) by the Zwicky Transient Facility (ZTF) survey using the Palomar Schmidt 48-inch (P48) Samuel Oschin telescope \citep{bellm, graham}. It was reported on the Transient Name Server (TNS) \citep{nordin} within 2\,hr of discovery. The first clear detection reported by the ZTF was in the $r$ band at a magnitude of $16.01 \pm 0.04$. The last nondetection with ZTF was 4 days (on 2020-05-02) before the discovery, with a reported upper limit of $g > 19.7$ mag. SN~2020jfo was classified as a Type II SN (\citealt{perley}) based on spectra taken with the Liverpool Telescope (LT) and the Nordic Optical Telescope (NOT). These spectra were taken $\sim 17$\,hr after the first ZTF detection. Our first spectrum with the Gran Telescopio Canarias (GTC) was taken nearly half an hour earlier than the LT and NOT spectra. The independent discoveries of this transient by ATLAS, PS2, {\it Gaia}, and MASTER were also reported to the TNS. We note here that SN~2020jfo was earlier studied by \citet{sollerman} and \citet{teja}.

SN~2020jfo exploded in the outskirts of the face-on spiral galaxy M61 at a redshift $z = 0.00522$ (from NED). M61 is a prolific SN-producing galaxy which has hosted seven other SNe to date. In the work by \citet{bose14}, the expanding photosphere method (EPM) was employed using the optical dataset from SN~2008in to estimate the distance ($14.51 \pm 1.38$ Mpc) calibrated with the atmosphere model provided by \citet{Dessart05}. We adopt this value of the distance for our work ($\mu = 30.81 \pm 0.20$ mag). \citet{sollerman} used this same distance for their work, whereas \citet{teja} used a distance of $16.45 \pm 2.69$ Mpc for their analysis. These distance values are in agreement with ours within the uncertainties.

\begin{figure}
	\includegraphics[width=\columnwidth]{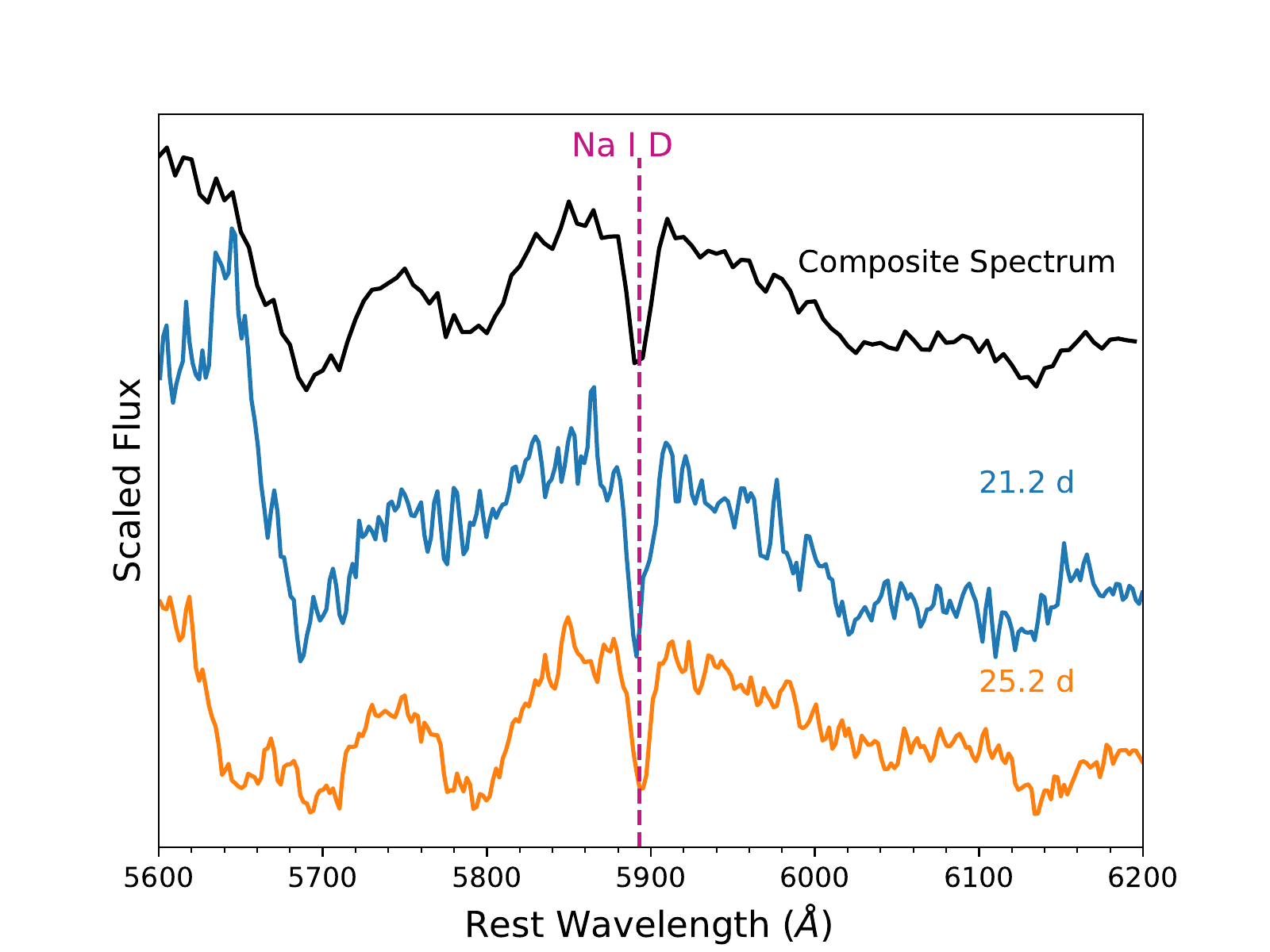}
    \caption{The narrow Na~I~D absorption dip due to the host galaxy, M61, in the two highest-SNR spectra (21.2 and 25.2 d) and the composite spectrum of SN~2020jfo.}
    \label{fig:extinction}
\end{figure}

The Galactic reddening along the line of sight to SN~2020jfo, obtained from the infrared (IR) dust maps given by \citet{schlafly}, is $E(B-V)_{\rm Gal} =  0.0194 \pm 0.0001$ mag. To estimate the host-galaxy extinction, empirical relations correlating the equivalent width (EW) of the Na~I~D doublet ($\lambda\lambda 5890$, 5896) and the colour excess are often used. The rest-frame spectra of SN~2020jfo do exhibit a conspicuous narrow absorption feature around 5893\,\AA. The EW of Na~I~D is estimated to be $0.88 \pm 0.21$\,\AA\ using the composite spectrum, constructed from the two highest signal-to-noise-ratio (SNR) spectra (21.2 and 25.2 d) of SN~2020jfo (see Figure \ref{fig:extinction}). Using the \citet{poznanski} empirical relations between EW and $E(B-V)$, we obtained $E(B-V)_{\rm host} = 0.15 \pm 0.09$ mag. The host-galaxy extinction can also be estimated from the colour \citep{olivares}, but this method is not very reliable \citep{dejaeger2018}. Using this method, we obtained negative values of $A_V$, which are unphysical. Since Na~I~D is a well-known tracer of gas in galaxies, and Na~I~D absorption is prominent in the spectra, we use the host-galaxy extinction estimated from the EW (although the reliability of this method has also been questioned; e.g., see \citealt{Phillips2013}). We adopt a total $E(B-V) = 0.17 \pm 0.09$ mag throughout this paper. \citet{sollerman} correct the photometric magnitudes only for the Galactic extinction, and \citet{teja} use a total $E(B-V) = 0.29 \pm 0.08$ mag. In Table \ref{tab:sn2020jfo}, we list the basic information on SN~2020jfo and its host galaxy.

\begin{table}
	\centering
	\caption{Basic information on SN 2020jfo and the host galaxy M61.$^a$}
	\begin{tabular}{ll} 
		\hline
		Host galaxy & NGC 4303\\
		Galaxy type & SAB(rs) bc\\
		Redshift & $0.00522 \pm 0.00001$\\
		Major axis diameter & $6.5'$\\
		Minor axis diameter & $5.8'$\\
		Heliocentric velocity & $1566.12 \pm 2.10$ km s$^{-1}$\\
		\\
		Distance & $14.51 \pm 1.38$ Mpc$^b$\\
		Total extinction $E(B-V)$ & $0.17 \pm 0.09$ mag$^c$\\
		SN type &  IIP\\
		Date of discovery (MJD) & 58975.2\\
		Estimated date of explosion (MJD) & 58973.1$^c$\\
		\hline
	\end{tabular}
	\newline
	\noindent
	$^a$The host-galaxy parameters are taken from NED.\\
	\noindent
	$^b$\citet{bose14}.\\
	\noindent
	$^c$This paper.\\
	\label{tab:sn2020jfo}
\end{table}

\section{Observations and Data Reduction}
\label{observations}

\subsection{Photometric observations}
\subsubsection{\textit{Swift}/UVOT photometry}
\label{sec:UVOT}

Photometry of SN~2020jfo was obtained with the Neil Gehrels \textit{Swift} Observatory (hereafter {\it Swift}) between 2020-05-07 and 2021-01-30 at 26 epochs. The transient was bright in all near-UV (NUV) and optical filters of the \textit{Swift}/UVOT instrument. The UVOT data were reduced using the standard pipeline available in the \texttt{HEAsoft} software package\footnote{\url{https://heasarc.nasa.gov/lheasoft/}}. Observations at each epoch were conducted using one or several orbits. To improve the SNR in a given band at a particular epoch, we coadded all orbit data for that corresponding epoch using the \texttt{HEAsoft} routine \texttt{uvotimsum}. We used the routine \texttt{uvotdetect} to determine the correct position of the transient (which is consistent with the ground-based optical observations) and the routine \texttt{uvotsource} to measure the apparent magnitude of the transient by performing aperture photometry. For source extraction we used a small aperture of radius 3\farcs5, while an aperture of radius $100''$ was used to determine the background. There is a possibility that when multiple photons are coincident on the same area of the UVOT detector, it does not count all the photons, an effect known as ``coincidence loss." The \texttt{uvotsource} routine makes the necessary correction to compensate for the coincidence loss. All of the UVOT photometry presented here is corrected for coincidence loss.

The field was also visited by \textit{Swift} in the year 2008, to follow the Type II SN~2008in \citep{roy}, which exploded at the outskirt of another arm of the host. We have used those previous UVOT frames to calculate the host contribution at the location of SN~2020jfo. The host magnitudes at the transient location are $19.38 \pm 0.07$, $19.34 \pm 0.08$, $19.02 \pm 0.05$, $18.49 \pm 0.04$, $17.61 \pm 0.07$, and $17.34 \pm 0.09$ mag in the UVOT $uw2$, $um2$, $uw1$, $uvu$, $uvb$, and $uvv$ bands, respectively. To obtain the host-subtracted magnitudes of SN~2020jfo, we first subtracted the coincidence-loss-corrected count rate of the galaxy computed from the pre-SN observations, from the coincidence-loss-corrected count rate of SN~2020jfo calculated from the images where the SN exists. Then the host-subtracted count rates of the SN were converted to AB magnitudes \citep{oke1983}. The uncertainties in the final photometry were obtained by propagating the error associated with the host magnitude and the error in the ``unsubtracted magnitudes" in quadrature. Our photometry in AB mag is listed in Table~\ref{tab:phot_uvot}.

\subsubsection{Optical photometry}
Photometric follow-up observations of SN~2020jfo were carried out with telescopes in the Las Cumbres Observatory (LCO) under the Global Supernova Project (GSP) in $BVg'r'i'$ filters between 2020-05-07 to 2021-07-13 at 53 epochs. Data between 2020-05-07 to 2021-02-22 at 44 epochs in the clear and $BVRI$ filters were also taken with the 1.04\,m Sampurnanand Telescope (ST), the 1.3\,m Devasthal Fast Optical Telescope (DFOT), and the 3.6\,m Devasthal Optical Telescope (DOT) located in ARIES, Nainital, India, as well as with the 0.76\,m Katzman Automatic Imaging Telescope (KAIT; \citealt{Filippenko2001}) and the 1.0\,m Nickel telescope at Lick Observatory. The earliest $r$-band photometric point was taken with the 10.4\,m GTC. The instrument details of the respective telescopes are given in Table \ref{tab:instrument}. Preprocessing of the data, including bias and flat correction, was done using Python modules. The \texttt{Cosmicray} task in \texttt{IRAF}\footnote{IRAF stands for Image Reduction and Analysis Facility distributed by the National Optical Astronomy Observatories which is operated by the Association of Universities for research in Astronomy, Inc., under cooperative agreement with the U.S. National Science Foundation.} was used for the removal of cosmic rays. Multiple frames observed on the same night were median-combined in order to increase the SNR. We performed image subtraction on the cleaned observed frames to compute the true flux of the transient. For the ST and DFOT observations, we used an \texttt{IRAF} and \texttt{DAOPHOT II} \citep{stetson_1987} based routine developed by \citet{roy0}, with the template taken on 2011-01-04 (see \citealt{roy}). The instrumental magnitudes of the SN were estimated using \texttt{DAOPHOT II}. 

The LCO GSP data were reduced with the \texttt{lcogtsnpipe}\footnote{\url{https://github.com/svalenti/lcogtsnpipe}} pipeline (see \citealt{2016MNRAS.459.3939V}). For the LCO data, image subtraction was carried out using High Order Transform of PSF ANd Template Subtraction (HOTPANTS\footnote{\url{https://github.com/acbecker/hotpants}}). We used the images obtained on 2021-07-17, when the SN faded below the detection threshold of the 1\,m LCO telescopes, as the template. The instrumental SN magnitudes were calibrated using the magnitudes of a number of local standard stars in the SN field obtained from the AAVSO Photometric All-Sky Survey (APASS\footnote{\url{https://www.aavso.org/apass}}) catalogue, whereas the ST and DFOT photometric data were calibrated using the magnitudes of the field standards marked by \citet{roy}. The data from KAIT and the 1.0\,m Nickel telescope were reduced using a custom pipeline\footnote{\url{https://github.com/benstahl92/LOSSPhotPypeline}} detailed by \citet{Stahl2019}. An image-subtraction procedure was also applied in order to remove the host-galaxy light.
The Pan-STARRS1\footnote{\url{http://archive.stsci.edu/panstarrs/search.php}} catalog was used for calibration.
Apparent magnitudes were all measured in the KAIT4/Nickel2 natural system, then transformed back to
the standard system using local calibrators and colour terms for KAIT4 and Nickel2 \citep{Stahl2019}.
Figure \ref{fig:subtracted_image} shows the template-subtracted image of SN~2020jfo. The calibrated SN magnitudes are listed in Tables \ref{tab:lco_phot}, \ref{tab:aries_phot}, and \ref{tab:shane_phot}.

\subsection{Optical and near-infrared spectroscopy}
Spectroscopic observations of SN~2020jfo were conducted at 26 epochs between 2020-05-06 and 2021-05-04 using the 2.0\,m Faulkes Telescope North (FTN) and Faulkes Telescope South (FTS) of the LCO Network (again via the Global Supernova Project), the 3.0\,m Shane telescope at Lick Observatory, the 3.6\,m DOT (ADFOSC; \citealt{Omar2019}), and the 10.4\,m GTC. The details of the instruments used are given in Table \ref{tab:instrument}. The one-dimensional (1D) wavelength- and flux-calibrated spectra were extracted using the \texttt{floydsspec} pipeline\footnote{\url{https://github.com/svalenti/FLOYDSpipeline}} \citep{Valenti2014} for the LCO data. Spectroscopic reduction of the Shane, DOT, and GTC data were done using standard tasks in \textit{IRAF} which include preprocessing, extraction of the 1D spectrum, wavelength calibration, and flux calibration. The slit-loss corrections were accounted for by scaling the spectra to the photometry at the same or nearby epoch. We also corrected the spectra for the heliocentric redshift of the host galaxy. The log of spectroscopic observations is presented in Table \ref{tab:spec_log}.

Near-infrared (NIR) spectra of SN~2020jfo were obtained with the SpeX \citep{2003PASP..115..362R} spectrograph installed on the 3.0\,m NASA Infrared Telescope Facility (IRTF) at three epochs (2020-05-13, 2020-05-24, and 2020-06-15). The spectra were taken in both the PRISM and SXD mode with a slit size of $0\farcs5 \times 15''$. The spectra were taken using the classic ABBA technique, and were reduced utilising the \texttt{Spextool} software package \citep{2004PASP..116..362C}. The telluric absorption corrections were done using the \texttt{XTELLCOR} software. The log of NIR spectroscopic observations is given in Table~\ref{tab:spec_log}.

\section{Light curves of SN~2020jfo}
\label{lightcurves}

\subsection{Light-curve rise and explosion epoch}
\label{explosion_epoch}

\begin{figure}
\includegraphics[width=\columnwidth]{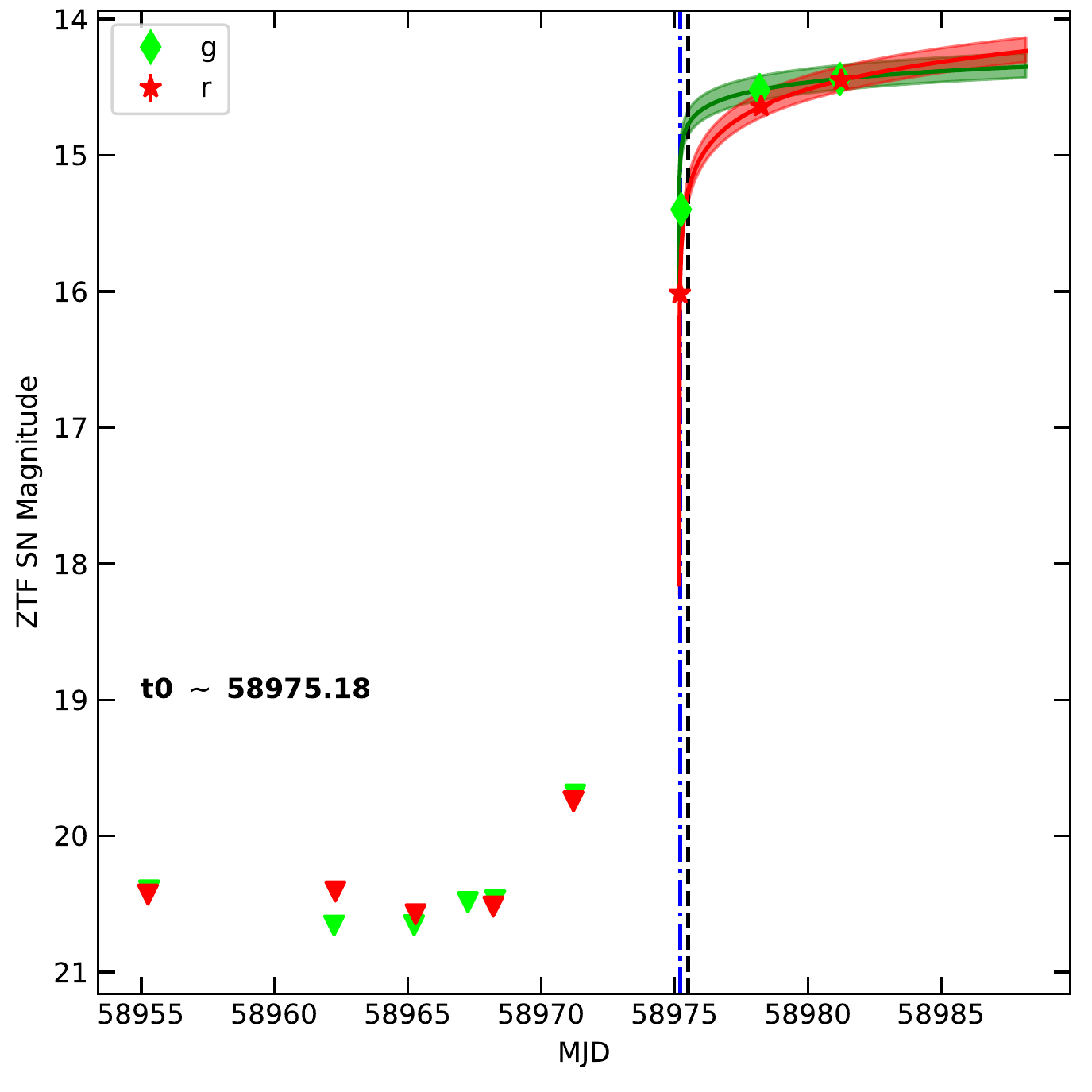}
\caption{The ZTF-public $g$-band and $r$-band data of the rising phase of SN~2020jfo. The downward arrows show the upper limits. The black dashed vertical line marks the epoch (MJD 58975.5) of the
first spectrum reported on the TNS. The blue dot-dashed vertical line marks
the date of discovery (MJD 58975.203), whereas the date of explosion 
($t_0$) obtained through power-law fitting is MJD 58975.18.} 
\label{fig:earlyLC}
\end{figure}

We used the ZTF-public data to compute the rise time of the transient. The object was first detected before its peak in the ZTF $r$ band on MJD 58975.203\footnote{\url{https://alerce.online/object/ZTF20aaynrrh}}, considered the epoch of discovery in this paper. A third-order polynomial fit to the early part of the ZTF $r$ and $g$ light curves shows that the transient attained its peak at MJD $58981.698 \pm 1.514$ in the observer's frame. Figure \ref{fig:earlyLC} represents the early-time ZTF photometry along with the pre-SN nondetection ($3\sigma$) limits. The limit in nondetection (difference between last nondetection and first detection) is roughly 3.9 days, which also indicates the early discovery of the SN. The epoch of spectral classification (MJD 58975.5) is marked by the black dashed vertical line, whereas the epoch of discovery is marked by the blue dot-dashed line. 
We have assumed that during the early rising phase, the transient's flux ($f$) evolved with time ($t \ge t_0$) according to a power law, $f\propto(t-t_0)^{\beta}$ ($f=0$ for $t \le t_0$). Here $t_0$ is the epoch of explosion and $\beta$ is the temporal index. The typical upper limits of the ZTF photometry during early nondetections are respectively $20.49 \pm 0.10$ mag and $20.47 \pm 0.10$ mag in $g$ and $r$. We have further assumed that for $t < t_0$, the fluxes at the SN location correspond to these upper limits. The power-law fit to the rising part of the light curves gives $t_0 \approx 58975.18$, consistent with the date of explosion computed by \cite{sollerman} and \cite{teja}.

A comparison of the classification spectrum taken with the Liverpool Telescope/SPRAT on 2020-05-06 (MJD 58975.92)\footnote{\url{https://www.wis-tns.org/object/2020jfo}} with early-time spectra of Type II events using \texttt{GELATO}\footnote{\url{https://gelato.tng.iac.es/}} shows that SN~2020jfo was roughly 3.8 days old on 2020-05-06. This implies that the epoch of explosion according to the spectral classification is MJD 58972.12. Furthermore, we run the SED matching code \texttt{SNID} \citep{2007ApJ...666.1024B} to constrain the explosion epoch of SN~2020jfo. We perform the spectral fitting of the same classification spectrum obtained with the Liverpool Telescope/SPRAT. The ``rlap" parameter quantifies the quality of the fit, with a higher value implying a better correlation. The best match, with rlap parameter 4.9, is found with the spectrum of SN~2013ej obtained 3.7 days after explosion. Based on this, we estimate the age of SN~2020jfo on 2020-05-06 to be 3.7 days since explosion, which translates to MJD 58972.20 as the date of explosion. \cite{anderson14} and \cite{gutirrez} have indicated that this method is consistent with the last-nondetection method with an uncertainty of $\pm 5$ days.

\begin{figure*}
\includegraphics[width=\textwidth]{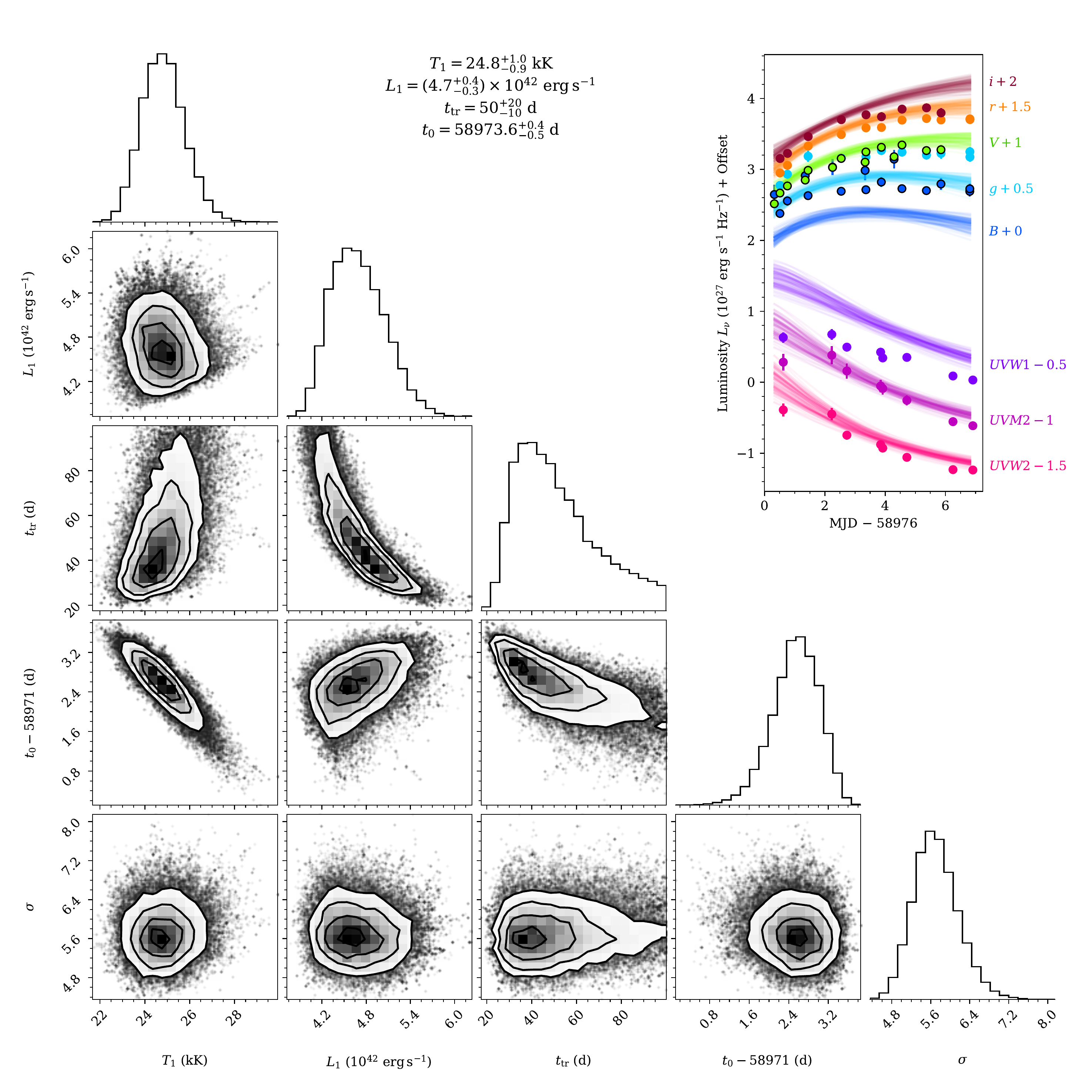}
\caption{The fit to the \citet{sapirwax} shock-cooling model using the method described by \citet{hossei2018}. Posterior probability distributions and correlation between parameters of SN~2020jfo are shown in the corner plot. The parameters corresponding to the temperature and luminosity of the SN 1 day after the explosion ($T_{1}$, $L_{1}$), the time of explosion ($t_{0}$), and the time at which the envelope becomes transparent ($t_{\rm tr}$) are shown. The top-right panel displays 100 fits randomly drawn from the MCMC routine \citep{griffin2020} to fit the early-time photometry. The $\sigma$ parameter depicts the intrinsic scatter of the model. The fit deviates largely from the observed light curves, and hence the estimated parameters from the fit are not used in the paper.}
\label{shock}
\end{figure*}

Additionally, we adopt early light-curve modeling, following the prescription of \citet{sapirwax}, to estimate the explosion epoch. The early-time emission from SNe is powered by shock cooling and provides useful insights on the radius and pre-explosion evolution of the progenitor. The Markov Chain Monte Carlo (MCMC) code developed by \citet{griffin2020} is used to obtain posterior probability distributions of temperature and luminosity of the SN 1 day after explosion ($T_{1}$, $L_{1}$), the time of explosion ($t_{0}$), and the time at which the envelope becomes transparent ($t_{\rm tr}$) simultaneously. We model the early-time light curve of SN~2020jfo using the multiband data up to MJD 58983.0 with $n=1.5$ for an RSG and uniform priors for all parameters. The model outputs the blackbody temperature and total luminosity as a function of time for each combination of the parameters, which are then translated to measured fluxes for each photometric point and all the light curves are simultaneously fitted. The parameter $\sigma$ is a measure of the intrinsic scatter of the model for which a log uniform prior is used. The posterior probability distributions and correlation between parameters of SN~2020jfo are shown in Figure~\ref{shock}. The model fits the {\it V}, {\it r}, and {\it i} bands reasonably well at least up to 58979 MJD. However, in the {\it B}, {\it g}, and {\it UV} bands, the model does not fit well; specifically, the {\it B} and {\it g} data are much redder and the {\it UVW1} band data are bluer than the models allow. Also, the model has a large intrinsic scatter ($\sigma = 5.7$). Thus, although the fit converges to an explosion epoch MJD $58973.6^{+0.4}_{-0.5}$ d, and the temperature and luminosity 1 day after explosion respectively converge to $T_{1} = 24.8^{+1.0}_{-0.9} \times 10^3$\,K and $L_{1} = 4.7^{+0.4}_{-0.3} \times 10^{42}$ erg~s$^{-1}$, we cannot accept these numbers as reliable estimates owing to the poor quality of the fits. The poor fitting of the model suggests that there is likely some circumstellar material (CSM) interaction which is powering the early-time light curve.

The explosion epoch from the different methods described above are 58975.0 (from the light curve rise), 58972.12 (from \texttt{GELATO}), and 58972.20 (from \texttt{SNID}). Here we use the mean MJD 58973.1 as the explosion epoch and the standard deviation of these estimates (1.6 day) as the uncertainty for further analysis, consistent with the explosion epochs MJD $58974.70^{+0.5}_{-3.5}$ and $58973.8^{+0.49}_{-0.46}$ used by \citet{sollerman} and \citet{teja}, respectively.

\subsection{Light-curve evolution of SN~2020jfo}
\label{sec:lc_analysis}

\begin{figure*}
	\includegraphics[width=\textwidth]{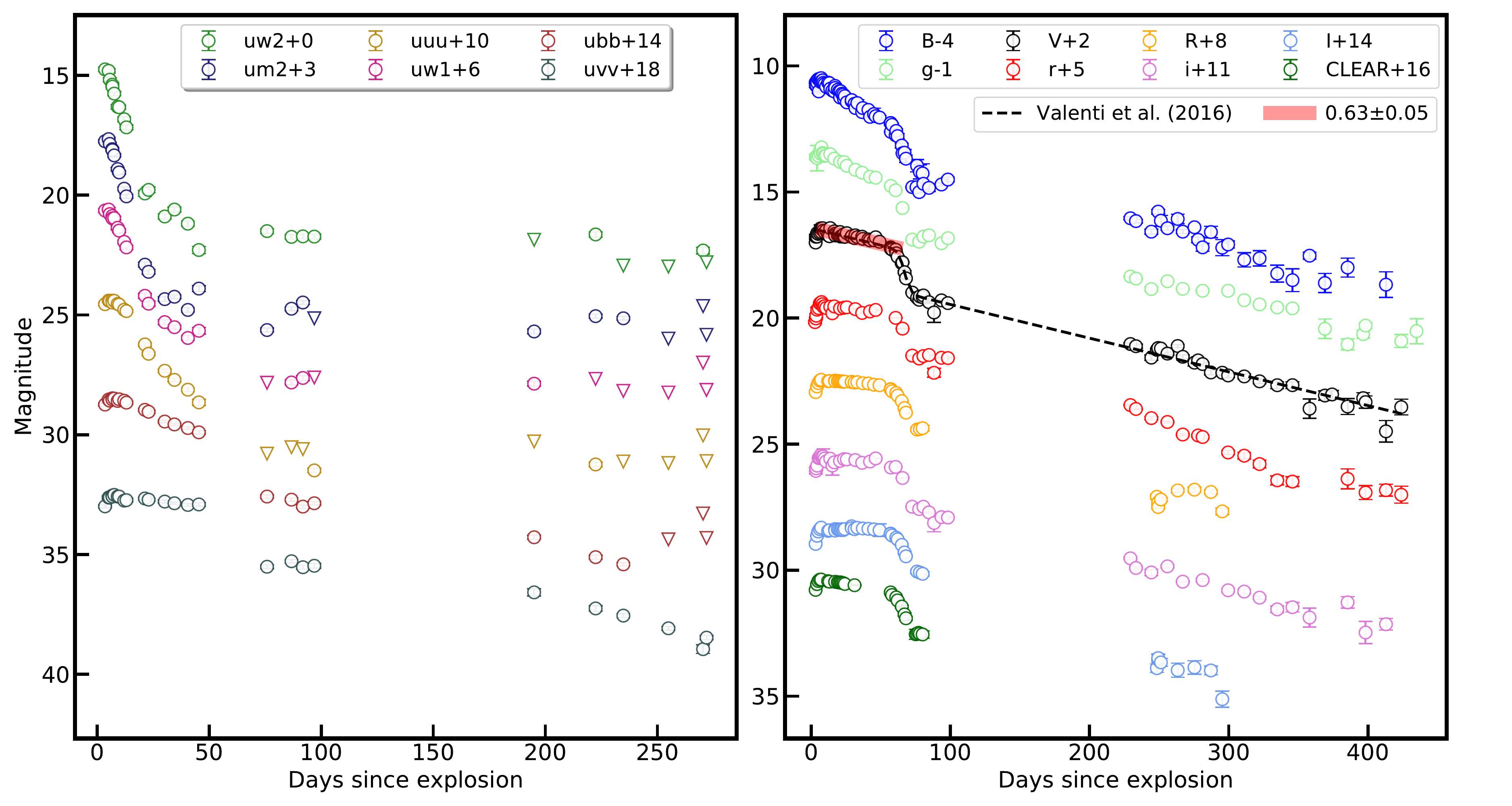}
    \caption{The {\it Swift} UVOT and optical light curves of SN~2020jfo since the date of explosion. The inverted triangles represent upper limits in the UVOT light curves. The light curves are offset by arbitrary numbers for display clarity. The dashed line represents the best fit to the $V$ band using \citet{valenti16}.}
    \label{lightcurve}
\end{figure*}

The {\it Swift} UVOT and multiband optical light curves of SN~2020jfo are shown in Figure \ref{lightcurve}. The optical light curves of SN~2020jfo initially rise to a peak in the {\it BVg'r'i'} bands. We estimate the peak magnitudes and the corresponding epoch at maximum brightness in the {\it BVg'r'i'} bands by fitting a parabolic function to the early-time data points ($t \leq 16$ day). The uncertainty in the maximum epoch is obtained by adding in quadrature the fitting error and the uncertainty in the explosion epoch. The values are listed in Table~\ref{peak_mag}. During the plateau phase, the SN declines with a rate of  $2.12 \pm 0.06$, $0.70 \pm 0.07$, and $0.30 \pm 0.08$ mag (50 day)$^{-1}$ in {\it B}, {\it V}, and {\it r}, respectively. As the SN evolves, the light curve falls from maximum and settles onto a plateau phase. The end of the plateau phase is marked by a sharp drop in magnitude, followed by an exponentially decaying phase powered by the radioactive decay of $^{56}$Co to $^{56}$Fe. The slope of the radioactive tail in $V$ is $1.60 \pm 0.07$ mag (100 d)$^{-1}$, which is steeper than the $^{56}$Co to $^{56}$Fe decay rate (0.98 mag (100 d)$^{-1}$), indicating partial trapping of $\gamma$-rays in the ejecta.

\begin{table}
\begin{center}
\caption{Peak magnitudes and phase of peak time in {\it BVg'r'i'} bands}
\label{peak_mag}
\begin{tabular}[width=\columnwidth]{c|c|c} 
	\hline
	Band & Epoch at maximum$^a$ & Peak magnitude\\
	     & ( days)         & (mag)\\
	\hline
	{\it B}  &	7.8$\pm$1.6 & 14.634$\pm$0.004\\
	{\it g'} &	8.1$\pm$1.6 & 14.484$\pm$0.001\\
	{\it V}  &	10.5$\pm$1.8 & 14.512$\pm$0.014\\
    {\it r'} &	10.2$\pm$1.6 & 14.460$\pm$0.008\\
    {\it i'} & 10.8$\pm$1.6 & 14.514$\pm$0.008\\
\hline
\end{tabular}
\newline
$^a$Phase with respect to the explosion epoch (MJD 58973.1).
\end{center}
\end{table}

We fit the {\it V}-band light curve using the expression given by \citet{valenti16},

\begin{center}
$y(t) = \frac{-a_0}{1\, + \, {\rm exp}{(t-t_{\rm pt})/w_0}} + p_0\,t + m_0$\,.
\end{center}

\noindent This is used to estimate the drop in magnitude from plateau to the tail phase, $a_0$, the duration of the transition phase ($\sim  6 \times w_0$), $t_{\rm pt}$, and a transition point between the end of the plateau (and the onset of the radioactive tail) and the decline rate of the radioactive tail phase, $p_0$. The estimated value of $t_{\rm pt}$ of SN~2020jfo is $67.0 \pm 0.6$ days, which is similar to that of SN 2006Y ($66.8 \pm 4$ days) but shorter than that of the other SNe~II in the comparison sample. \cite{teja} estimated a plateau length of $58 \pm 2$ days, which corresponds to the parameter OPTd from \cite{anderson14}, whereas we estimated $t_{\rm pt}$, which is roughly equivalent to OPTd + 0.5 times the duration of the transition phase. Considering this, our estimated $t_{pt}$ value is consistent with \cite{teja} within the uncertainties. The best-fitting value of $a_0$ is $1.68 \pm 0.09$ mag, which lies in the range of Type II SNe (1--2.6 mag) as suggested by \cite{valenti16}. SN II 2006Y has also been reported to have a similar drop in magnitude ($1.75^{+0.23}_{-0.15}$) during the transition phase. However, the duration of the transition phase ($6 \times w_0$) of SN~2020jfo (13.3 days) is shorter than that of SN II 2006Y (25.7 days). This could be due to different amounts of Ni mixing \citep{goldberg}. The values of best-fit parameters are listed in Table~\ref{plateau} and the best fit in the $V$ band is shown in Figure~\ref{lightcurve} by the dashed line.

\begin{table}
\caption{The best-fitted s$_{50V}$ slope of {\it V}-band lightcurve of SN 2020jfo between start and end phase. The measurement of slope is given in magnitudes per 50 days. The best fitted {\it V}-band light curve parameters using \citet{valenti16} model are also given here. }
\begin{tabular}[width=\columnwidth]{cccc} 
	\hline
	Slope & Decline Rate &  Start Phase & End Phase\\
	\\
     s$_{50V}$ & 0.63$\pm$0.05 & 8.8 & 60.7 \\ 
	\hline
	t$_{pt}$ (d) & a$_0$ (mag) &  w$_0$ (d) & p$_0$ (mag/d)\\
	\\
	67.0 $\pm$ 0.6 & 1.68 $\pm$ 0.09 & 2.22 $\pm$ 0.68 & 0.0134 $\pm$ 0.0003 \\ 
	\hline
\end{tabular}
\label{plateau}
\end{table}

\begin{table*}
\centering
\caption{Properties of the SNe in the comparison sample}
\label{tab:comparison_objects}
\begin{threeparttable}
\begin{tabular}{llccccccl}
\hline
Object   & Host Galaxy  & Distance  & Total & Ejected & M$_V^{peak}$& M$_V^{50d}$  & t$_{pt}$  & References\\
& & & Extinction  & $^{56}$Ni Mass  & & &   &\\
& & (Mpc) & {\it E(B-V)} (mag) & (\(M_\odot\)) & (mag) & (mag) & (days) &\\
\hline
SN~1995ad & NGC2139 & 22.91$\pm$1.64 & 0.035 & 0.028 & -17.64$\pm$0.29 & -16.27$\pm$0.11 & 50$^\dagger$ & 6 \\
SN~1999em & NGC 1637 & 11.7$\pm$0.1 & 0.10 & 0.042 & -16.86$\pm$0.07 & -16.71$\pm$0.06 & 118.1$\pm$1.0 & 1, 2, 3, 7 \\
SN~2005cs & M 51 & 8.40 & 0.12 & <0.003  & -14.81$\pm$0.37 & -14.55$\pm$0.32 & 126.0$\pm$0.5 & 4, 5, 7\\

SN~2006Y & anonymous & 146.0 & 0.11 & >0.050 & -18.09$\pm$0.01 & -16.75$\pm$0.02 & 66.8$\pm$4 & 7, 8\\
SN~2006ai & ESO 5-G9 & 67.5 & 0.11 & >0.034 & -18.23$\pm$0.02 & -17.00$\pm$0.01 & 72.4$\pm$5 & 7, 8\\
SN~2007od & UGC12846 & 25.7$\pm$1.84 & 0.038 & 0.02 & -18.00$\pm$0.16 & -17.31$\pm$0.13 & 45$^\dagger$ & 9\\
SN~2015ba & IC 1029 & 34.8$\pm$0.7 & 0.46$\pm$0.05 & 0.032$\pm$0.006 & -17.68$\pm$0.17 & -17.05$\pm$0.15 & 140.7$\pm$0.2 & 10\\
SN~2016egz & anonymous & 100.0 & 0.01 & -- & -18.46$\pm$0.03 & -17.55$\pm$0.05 & 71.2$^\ddagger$ $\pm$ 0.5 & 8\\
SN~2020jfo & M61 & 14.51$\pm$1.38 & 0.17$\pm$0.09 & 0.03$\pm$0.01 & -16.90$\pm$0.34 & -16.69$\pm$0.44 & 67.0$\pm$0.6 & This work\\
\hline
\end{tabular}
\begin{tablenotes}
\setlength\labelsep{0pt}
\footnotesize{\item $^{\dagger}$ OPTd
             \item $^{\ddagger}$ t$_{pt}$ estimated in this work
             \item (1) \citet{hamuy2001}, (2) \citet{leonard2002}, (3) \citet{leonard2003}, (4) \citet{Pastorello2006}, (5) \citet{Pastorello2009}, (6) \citet{Inserra1995ad}, (7) \citet{valenti16}, (8) \citet{hiramatsu}, (9) \citet{inserra2011}, (10) \citet{dastidar}}
\end{tablenotes}
\end{threeparttable}
 
\end{table*}

\begin{figure*}
	\includegraphics[width=\textwidth]{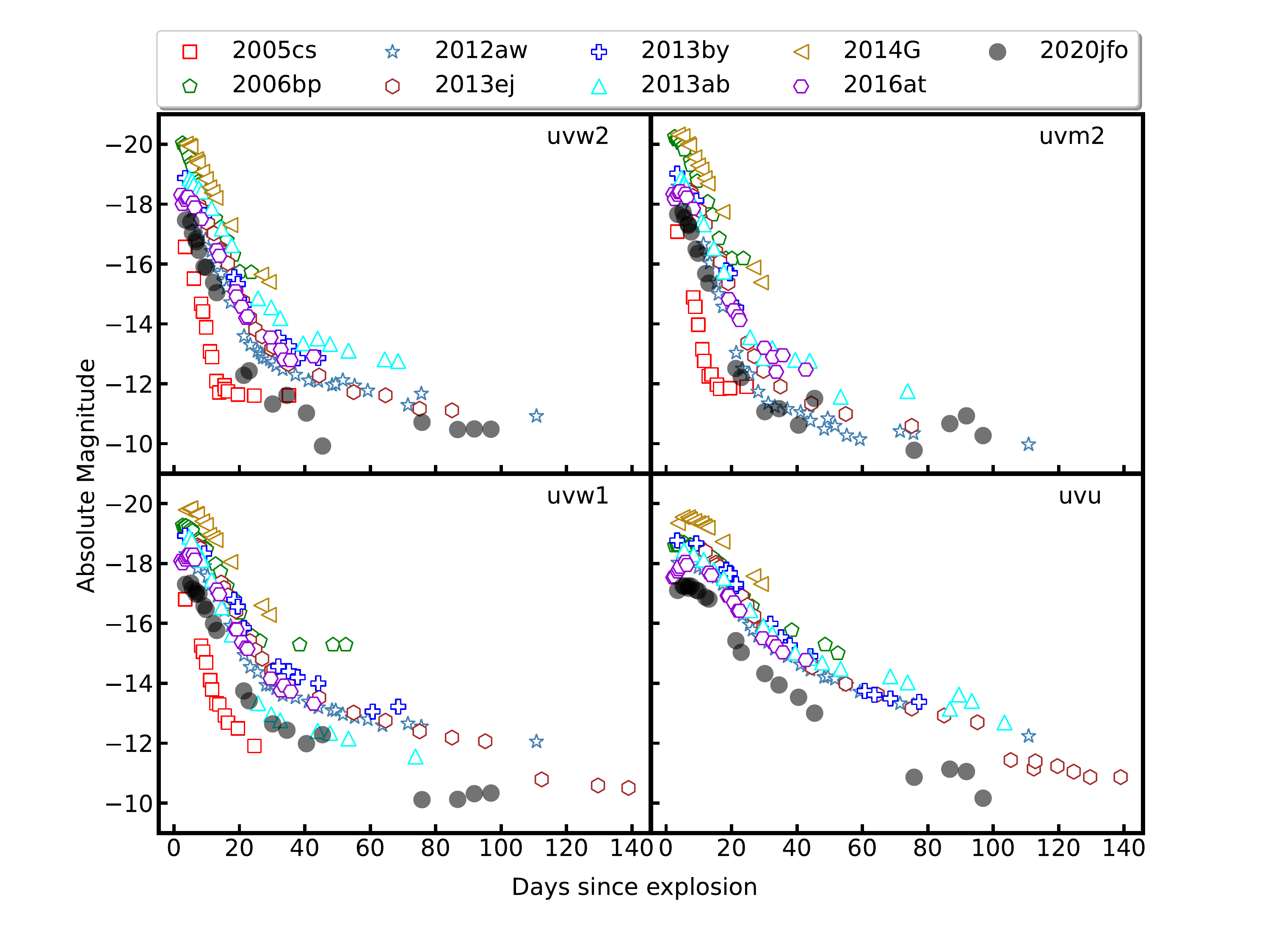}
    \caption{The comparison of \textit{Swift} UVOT absolute magnitude light curves of SN~2020jfo with those of some well-observed Type IIP SNe observed with \textit{Swift}. }
    \label{absolute_mag_uv}
\end{figure*}

SN~2020jfo has good UVOT data coverage until late times, unlike most other SNe~II, which have low apparent luminosity owing to host-galaxy extinction and large distances. The UVOT light curves decay rapidly until $\sim 23$ days since explosion with a rate of $0.28 \pm 0.01$, $0.30 \pm 0.01$, and $0.21 \pm 0.01$ mag day$^{-1}$ in the uw2, um2, and uw1 bands respectively. We compare the UV absolute magnitudes in the UVOT filters of SN~2020jfo with those of some other well-observed Type II SNe in Figure \ref{absolute_mag_uv}. The sample is chosen based on the availability of good cadence light curves in UV bands. This sample comprises the following SNe: 
SN 2005cs (\citealt{Pastorello2009, brown2009}), 
SN 2006bp (\citealt{dessert2006bp}), SN 2012aw (\citealt{bose2012aw}), 
SN 2013ab (\citealt{bose2013ab}), 
SN 2013by (\citealt{valenti2013by}),
SN 2013ej (\citealt{bose2013ej}),
SN 2014G (\citealt{bose2014G}),
and SN 2016at (\citealt{bose2016at}). All Type II SNe show a rapid decline in the first 20 days of evolution in the UV filters. SN~2020jfo falls on the fainter end of the normal-luminosity SNe~II and is brighter than the low-luminosity SN~2005cs. Some of the comparison SNe with good data coverage, such as SNe~2013by and 2013ab, show a constant-luminosity phase in the UV bands corresponding to the plateau phase in optical bands. However, the UV light curves of SN~2020jfo do not exhibit any plateau phase.

To examine the behaviour of SN~2020jfo in the optical bands, we assemble a comparison sample of SNe with varying plateau lengths and a range of explosion properties: 
SNe 1995ad (short plateau; \citealt{Inserra}), 1999em (normal Type IIP; \citealt{elmhamdi}), 
2005cs (low-luminosity Type IIP; \citealt{Pastorello2006}, \citealt{Pastorello2009}), 
2006Y (short plateau; \citealt{hiramatsu}),
2006ai (short plateau; \citealt{hiramatsu}), 2007od (short plateau; \citealt{inserra2011}),
2015ba (long plateau Type IIP; \citealt{dastidar}), and 
2016egz (short plateau; \citealt{hiramatsu}). The parameters of the comparison sample are listed in Table \ref {tab:comparison_objects}. 

\begin{figure}
	\includegraphics[width=\columnwidth]{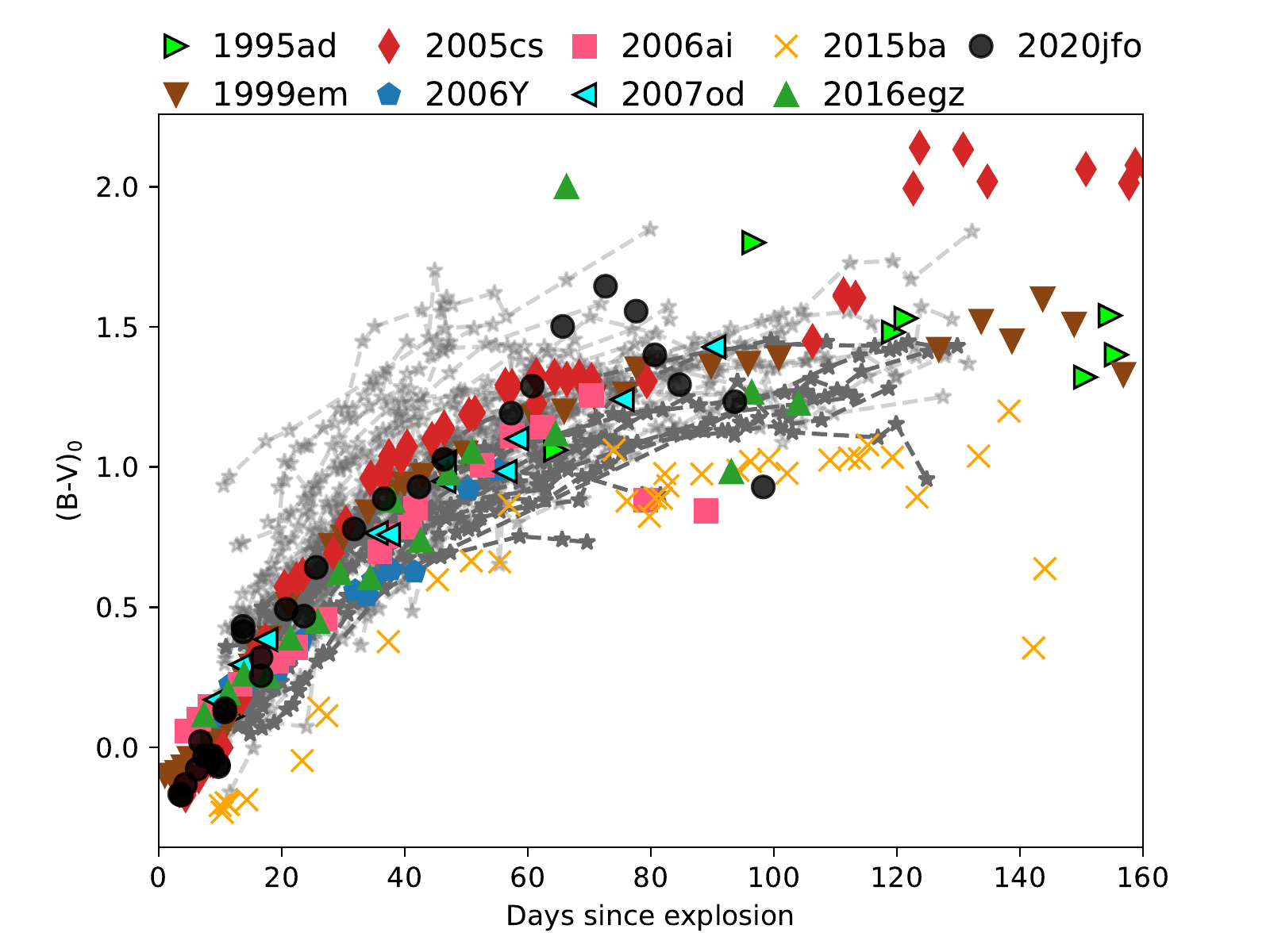}
    \caption{$(B-V)_0$ colour evolution of SN~2020jfo compared with the SN~II sample used in this paper.}
    \label{Color}
\end{figure}

The colour evolution provides information about the dynamics of the SN ejecta. The temporal evolution of the reddening-corrected $(B-V)_0$ colour of SN~2020jfo is depicted in Figure~\ref{Color} along with that of the comparison sample of SNe and the sample of Type II SNe from \citet{dejaeger2018} (shown in grey) and a subset of this sample with negligible host extinction (shown with grey stars). All SNe exhibit roughly the same trend in $(B-V)_0$ colour --- a gradual transition to redder colours in the first 50 days, followed by a flattening until the end of the plateau phase. Owing to expansion of the ejecta, the $(B-V)_0$ colour of SN~2020jfo becomes redder by 1.2 mag until the end of the plateau phase. Thereafter, SN~2020jfo starts transitioning from the plateau to the tail phase at $\sim 70$ days, which is marked by a steeper rise to redder colours. As the SN settles in the exponentially decaying tail phase, the $(B-V)_0$ colour gradually becomes bluer by 0.7 mag. SN~2020jfo and the other SNe~II with shorter plateau lengths (SNe 1995ad, 2006Y, 2006ai, 2007od, and 2016egz) do not show the constant-colour phase corresponding to the plateau phase as seen in other SNe~II such as SNe 1999em, 2005cs, and 2015ba; almost immediately after 50 days, the light curve transitions from the plateau to the tail phase in Type II SNe with shorter plateau. Like SN~2020jfo, a sudden transition to redder colours at the end of the plateau phase is also evident in SN~2005cs; however, the latter does not display a transition to bluer colours in the nebular phase. SN 2006ai also exhibits this transition to bluer colours in the nebular phase.

\begin{figure*}
	\includegraphics[width=\textwidth]{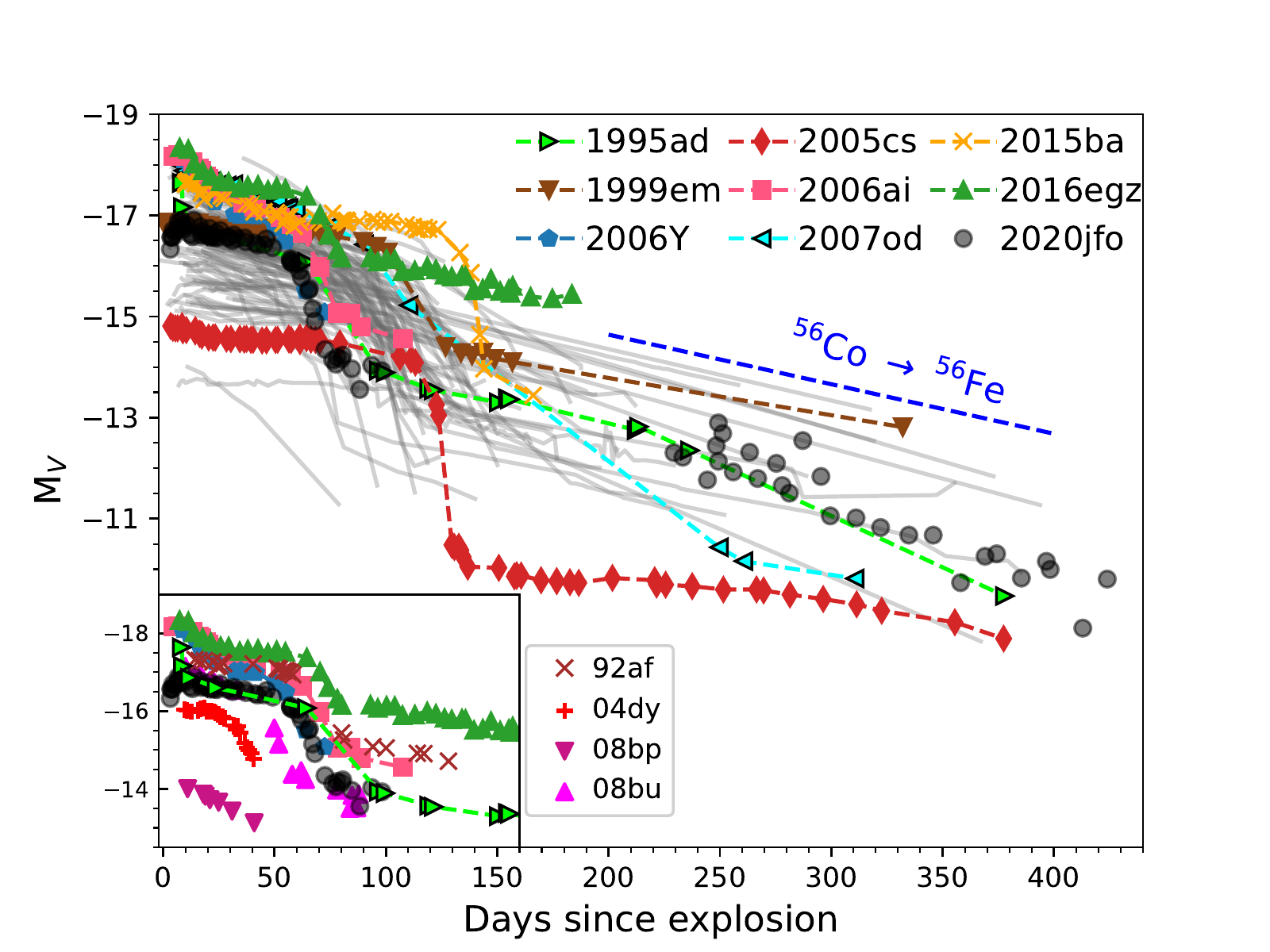}
    \caption{Absolute $V$-band light curve of SN~2020jfo compared with the Type II SNe in the comparison sample. The grey lines represent the light curves of 116 SNe~II having different plateau lengths from \protect\cite{anderson14}. The inset shows the comparison of SN~2020jfo with other Type II SNe having similar plateau lengths in the comparison sample used in the paper and the sample of \protect\cite{anderson14}.}
    \label{absolute_mag}
\end{figure*}

The absolute magnitude $V$-band light curve of SN~2020jfo is shown along with the comparison sample in Figure~\ref{absolute_mag}. In terms of brightness, SN~2020jfo has similar absolute magnitudes in the early and plateau phase as the prototypical Type IIP SN~1999em, with a remarkably early drop from the plateau as compared to SN~1999em. The plateau length of SN~2020jfo is rather similar to that of the Type II SNe 1995ad, 2006Y, and 2006ai with plateau duration in the range 50--70 days. The peak of SN~2020jfo is $-16.90 \pm 0.34$ mag, consistent within uncertainties with that obtained by \cite{teja}. The Type II SNe having shorter plateau duration (SNe 1995ad, 2006Y, 2006ai, 2007od, and 2016egz) are brighter than SN~2020jfo by at least 0.7 mag at peak and have a steeper post-peak decline rate (up to 20 days) than SN~2020jfo (2.5 mag (50 day)$^{-1}$, SN 1995ad; 3.0 mag (50 day)$^{-1}$, SN 2006Y; 1.7 mag (50 day)$^{-1}$, SN 2006ai; 1.2 mag (50 day)$^{-1}$, SN 2007od; 2.2 mag (50 day)$^{-1}$, SN 2016egz). SN~2020jfo is brighter than the low-luminosity Type II SN 2005cs by $\sim 2$ mag in the plateau phase and fainter than the long-plateau SN~2015ba by 0.5 mag. The grey lines represent the light curves of 116 Type II SNe having different plateau lengths from \cite{anderson14}. The sample of \cite{anderson14} also consists of a few SNe~II with relatively shorter plateaus (SNe~1992af, 2004dy, 2008bp, and 2008bu), which we have shown separately in the inset plot along with SNe~2020jfo, 1995ad, 2006Y, 2006ai, and 2016egz. Except for SNe 1992af, the other SNe~II (SNe 1995ad, 2004dy, 2008bp, and 2008bu) have absolute magnitudes during the plateau phase similar to or lower than that of SN~2020jfo. As noted by \cite{hiramatsu}, SN 1992af has an uncertain explosion epoch and hence might have a longer plateau duration. The tail-phase luminosity of SN~2020jfo is similar to that of SN~1999em and lower than that of the rest of the SNe in the comparison sample except for SN~2005cs. SN 2016egz has the highest tail luminosity, whereas SNe 2006Y and 2006ai have tail luminosities similar to that of SN~1999em. The inset plot shows that the Type II SNe with similar plateau magnitudes also have comparaable tail luminosities.

The $^{56}$Ni mass is estimated from the tail bolometric luminosity following \citet{hamuy2003} using the tail $V$-band magnitudes. The tail luminosity ($L_t$) obtained from the $V$ magnitudes using the bolometric correction factor given by \citet{hamuy2003} at two epochs (72.2 and 80.2 days) are $2.19 \pm 0.71 \times 10^{41}$ erg s$^{-1}$ and $2.04 \pm 0.66  \times 10^{41}$ erg s$^{-1}$, respectively, which corresponds to a mean $^{56}$Ni mass of $0.03 \pm 0.01$ M$_\odot$. This is consistent with the $^{56}$Ni mass obtained for SN~2020jfo by \cite{sollerman} and \cite{teja}.

\section{Spectral Analysis}
\label{spectroscopy}
\subsection{Spectroscopic evolution}

\begin{figure*}
	\includegraphics[width=\columnwidth]{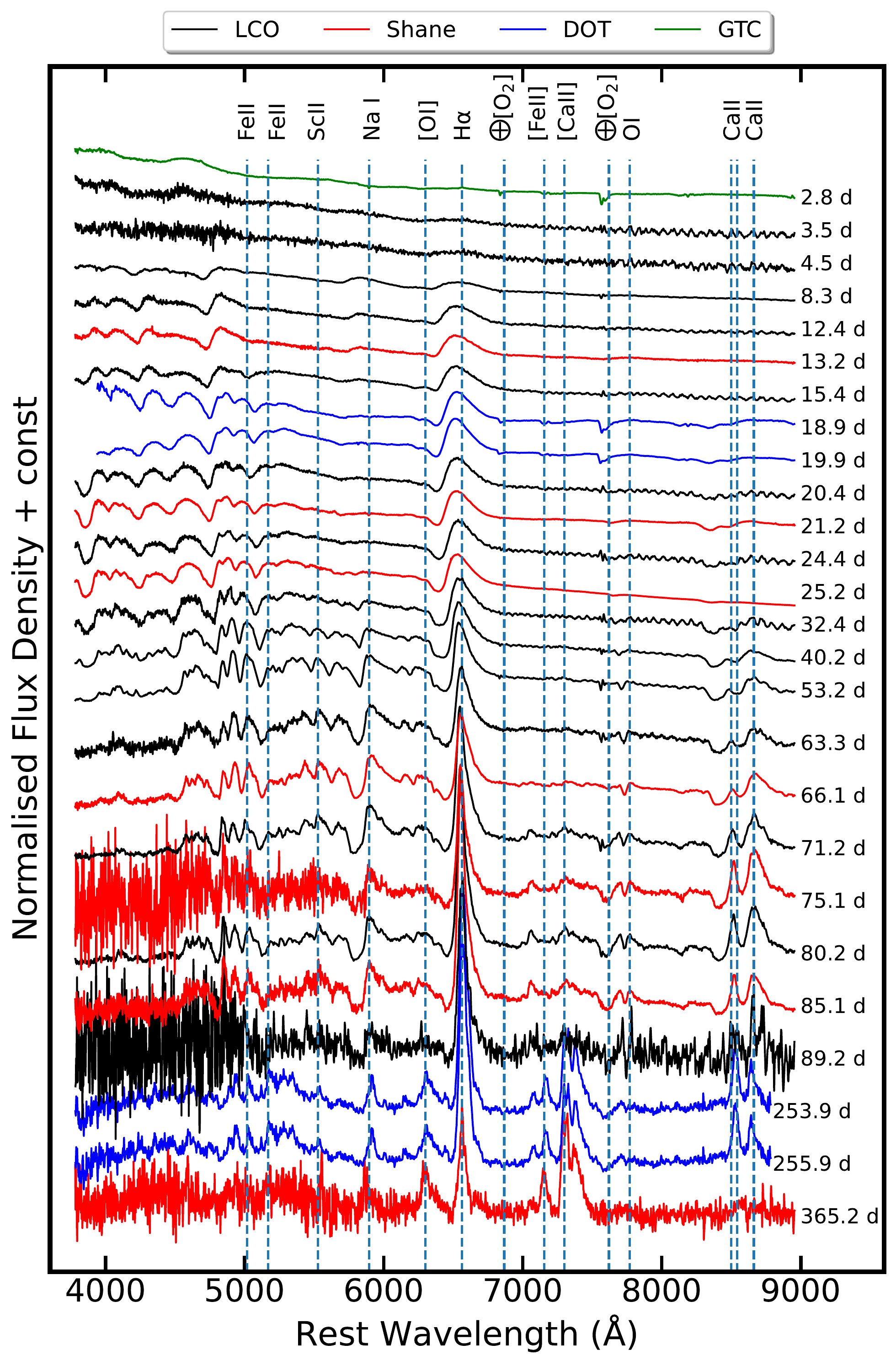}
    \caption{The spectral evolution of SN~2020jfo from 2.8 days to 365.2 days since explosion with the prominent spectral features marked. Each spectrum is normalised with the median flux.}
    \label{spectral_evolution}
\end{figure*}

The spectral evolution of SN~2020jfo from 2.8 to 365.2 days since explosion is shown in Figure \ref{spectral_evolution}. At the early phase of the evolution (2.8--4.5 days), broad P~Cygni profiles of H Balmer lines can be seen superimposed on a blue continuum. The spectra from 8.3 days reflect the transition from a hot to a cool SN envelope, as the photosphere begins to penetrate deeper into the metal-rich ejecta. These spectra mark the emergence of other lines from heavier atomic species such as Ca, Fe, Sc, Ba, Ti, and Na. Among these, the Fe~II $\lambda\lambda$4950, 5169 lines appear from 18.9 days. Sc~II lines, O~I, and the Ca~II NIR triplet gain prominence from day 40.2 and get stronger over time. By the end of the plateau phase, all weak and blended lines have evolved and become conspicuous. A high-velocity absorption dip blueward of the H$\alpha$ absorption minimum becomes apparent in the day 63--85 spectra.

\begin{figure*}
	\includegraphics[width=\textwidth]{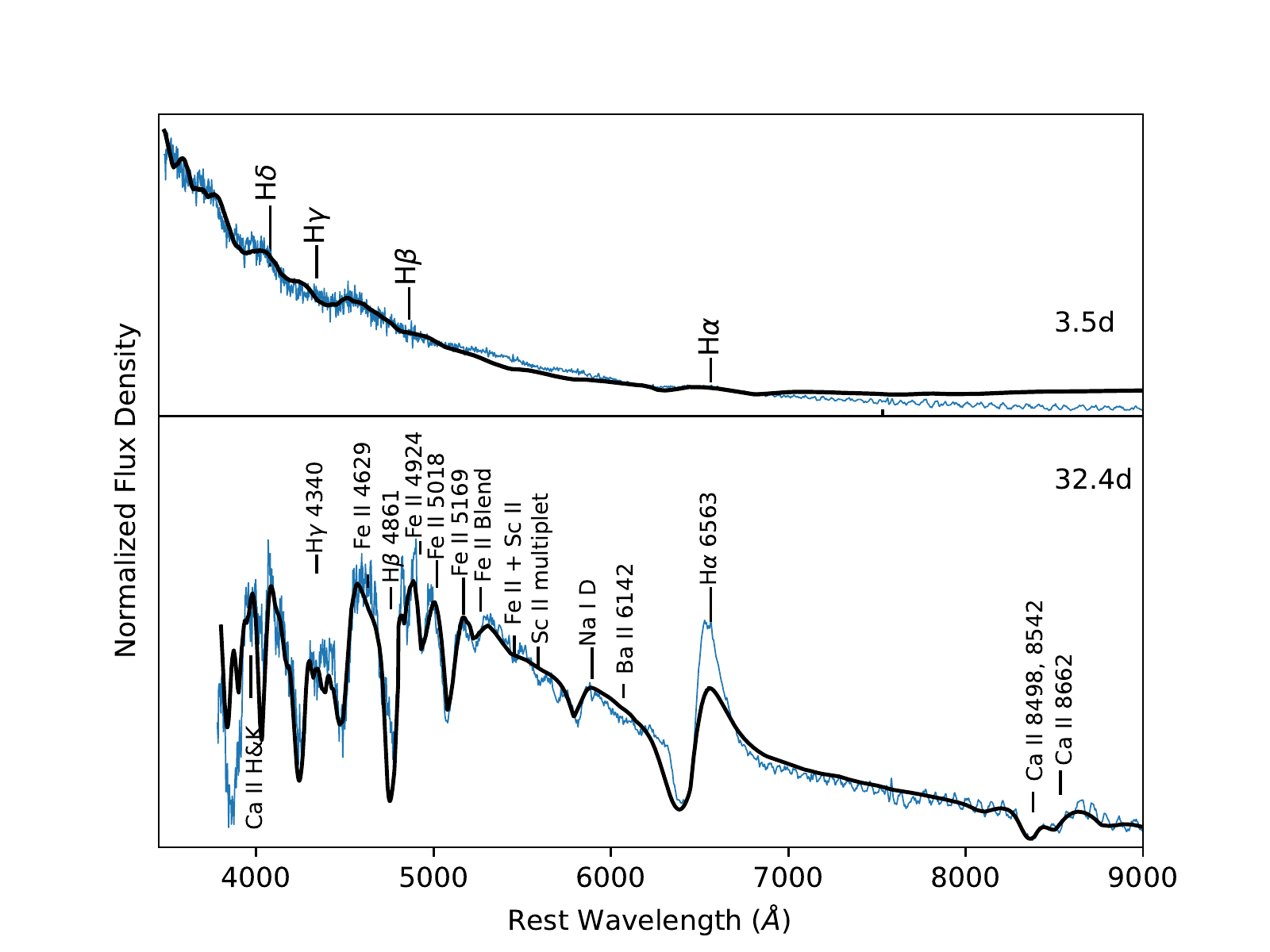}
    \caption{\texttt{SYNAPPS} modelling of the early (3.5 day) and plateau phase (32.4 day) spectra of SN~2020jfo (shown in black) marked with the lines identified from the spectral modelling.}
    \label{spectra_modelling}
\end{figure*}

For robust line identifications, we used \texttt{SYNAPPS} \citep{2013ascl.soft08007T} to reproduce the early-time (3.5 d) and photospheric-phase (32.4 d) spectra of SN~2020jfo (Figure \ref{spectra_modelling}). For modelling the early spectrum, we used H~I and Ca~II species. The photospheric velocity and blackbody temperature at 3.5 d as constrained from the modelling are 13,360 km s$^{-1}$ and 11,650 K, respectively. The photospheric spectrum is reproduced using H~I, Ca~II, Fe~II, Na~I, Ba~II, Sc~II, and Ti~II lines. The Ti~II, Sc~II, and Ba~II species mainly contribute in reproducing the 5200--6000\,\AA\ region. The photospheric velocity and blackbody temperature at 32.4 d as constrained from the modelling are 5476 km s$^{-1}$ and 8380 K, respectively. In both cases, the ``detach" parameter has been deactivated, which means that the minimum ejecta velocity of all the ions will be the same as the photospheric velocity.

The day 63.3 spectrum marks the onset of the transition from the plateau to the nebular phase. All subsequent spectra up to day 85.1 are representative of the early nebular phase, when the outer ejecta have become optically thin. In the late nebular phase spectra (253.9--365.2 day), in addition to the H$\alpha$ emission, a number of forbidden emission lines such as [O~I] $\lambda\lambda$6300, 6364 and [Ca~II] can be seen. Also, the nebular spectra of SN~2020jfo show strong [Ni~II] and [Fe~II] forbidden emission which can be used as a diagnostic of the Ni-to-Fe mass ratio in the entire ejecta \citep{Jerkstrand2015J}. 

\begin{figure}
	\includegraphics[scale=0.35, clip, trim={2.5cm 2.5cm 1cm 1cm}]{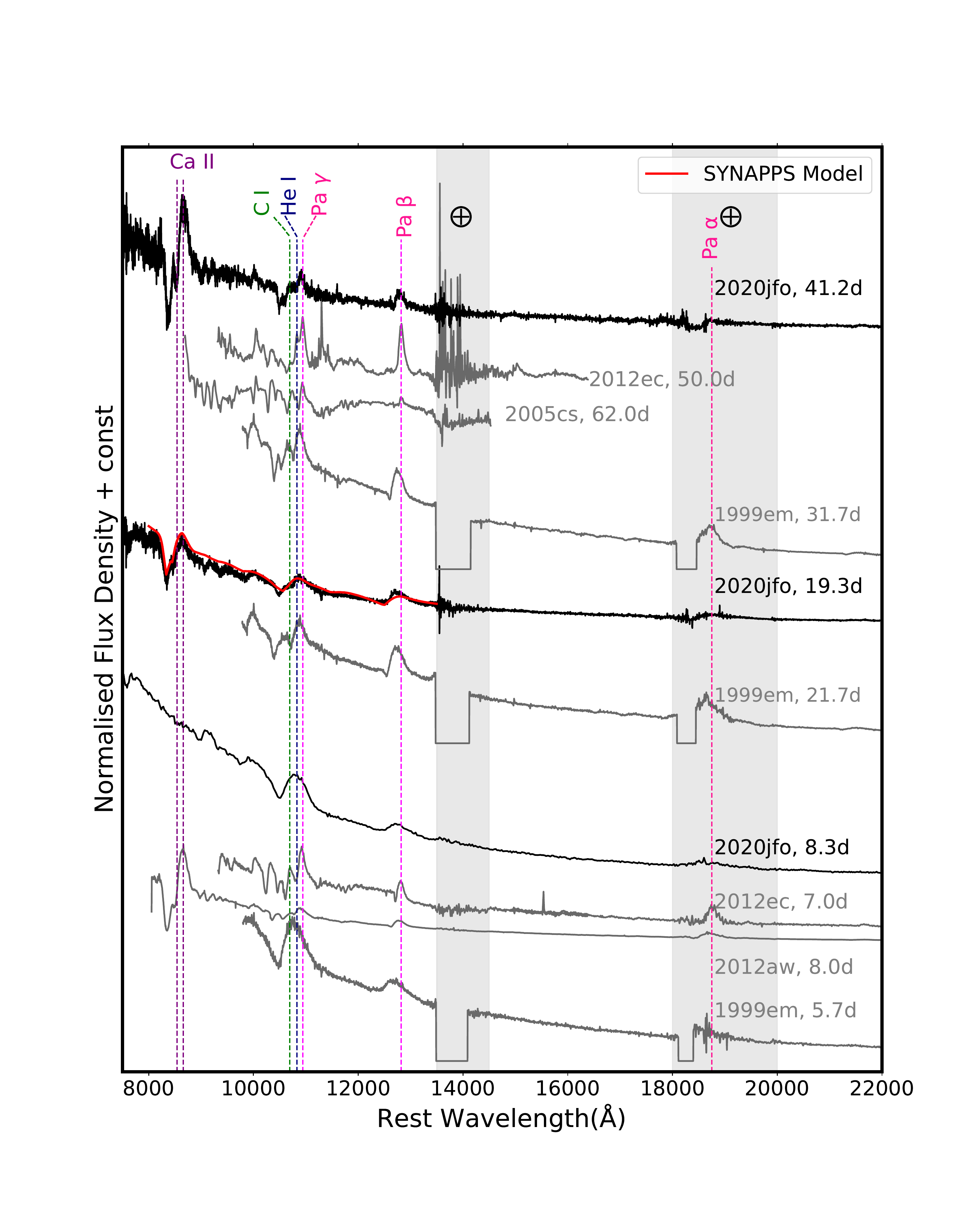}
    \caption{The NIR spectral evolution of SN~2020jfo compared with that of other Type II SNe. The \texttt{SYNAPPS} model of the day 19.3 spectrum in the wavelength range 8000--13,500\,\AA\ using H~I, He~I, C~I, and Ca~II species is shown in red.
    }
    \label{nirspec}
\end{figure}

The three NIR spectra of SN~2020jfo obtained on days 8.3, 19.3, and 41.2 are shown in Figure \ref{nirspec}.
The day 8.3 spectrum exhibits weak lines of Pa$\alpha$, Pa$\beta$, Pa$\gamma$, and He~I, and a relatively strong Ca~ II NIR triplet. The feature around $\lambda$10,800 in the day 8.3, 19.3, and 41.2 spectra is probably a He~I and Pa$\gamma$ blend, although there is a possibility of C~I formation at 10,691\,\AA. We used SYNAPPS to model the wavelength range 8000--13,500\,\AA\ in the day 19.3 spectrum using H~I, He~I, C~I, and Ca~II ions, and can reproduce this region well as shown in Figure \ref{nirspec}. The presence of the C~I line has been indicated in a few SNe~II, such as SNe1999em \citep{2001ApJ...558..615H}, 2005cs \citep{Pastorello2009}, and 2012aw \citep{2019ApJ...887....4D}. We compare the NIR spectra of SN 2020jfo with those of other SNe~II like SNe 1999em \citep{2001ApJ...558..615H}, 2005cs \citep{Pastorello2009}, 2012aw \citep{2014ApJ...787..139D}, and 2012ec \citep{2015MNRAS.448.2312B} at nearby epochs in Figure \ref{nirspec}. Overall, the NIR spectra of SN~2020jfo show similarity with other SNe~II. The day 8.3 spectrum of SN~2020jfo is similar to the day 5.7 spectrum of SN~1999em and the day 7.0 spectrum of SN~2012ec, all showing the classical feature at 10,830\,\AA, while the day 8.0 spectrum of SN~2012aw does not exhibit any conspicuous line at this wavelength. The 10,549\,\AA\ absorption feature in the day 21.7 spectrum of SN~1999em was suggested to be possibly a blend of C~I $\lambda$10,691 and He~I $\lambda$10,830 \citep{2001ApJ...558..615H}. However, in the day 62.0 spectrum of SN~2005cs, this feature was suggested to be mostly due to P$\gamma$, Sr~II $\lambda$10,915, and C~I $\lambda$10,691, with less dominance of He~I $\lambda$ 10,830 than in other CCSNe \citep{Pastorello2009}. In SN~2020jfo, this feature is mostly dominated by Pa$\gamma$, He~I, and C~I. There is possibly no formation of Sr~II at least until day 19.3 in SN~2020jfo; introducing the ion deteriorated the SYNAPPS fit.

\subsection{Comparison with other SNe}

We compare the spectra of SN~2020jfo at early times ($\sim 8.3$ day), the plateau ($\sim 40.2$ day), and the nebular phase ($\sim 365.2$ day) with the spectra of the comparison sample at similar epochs. The spectroscopic comparison sample is composed of the same SNe~II that are defined in Section \ref{sec:lc_analysis} and listed in Table \ref{tab:comparison_objects}.

\begin{figure}
	\includegraphics[width=\columnwidth]{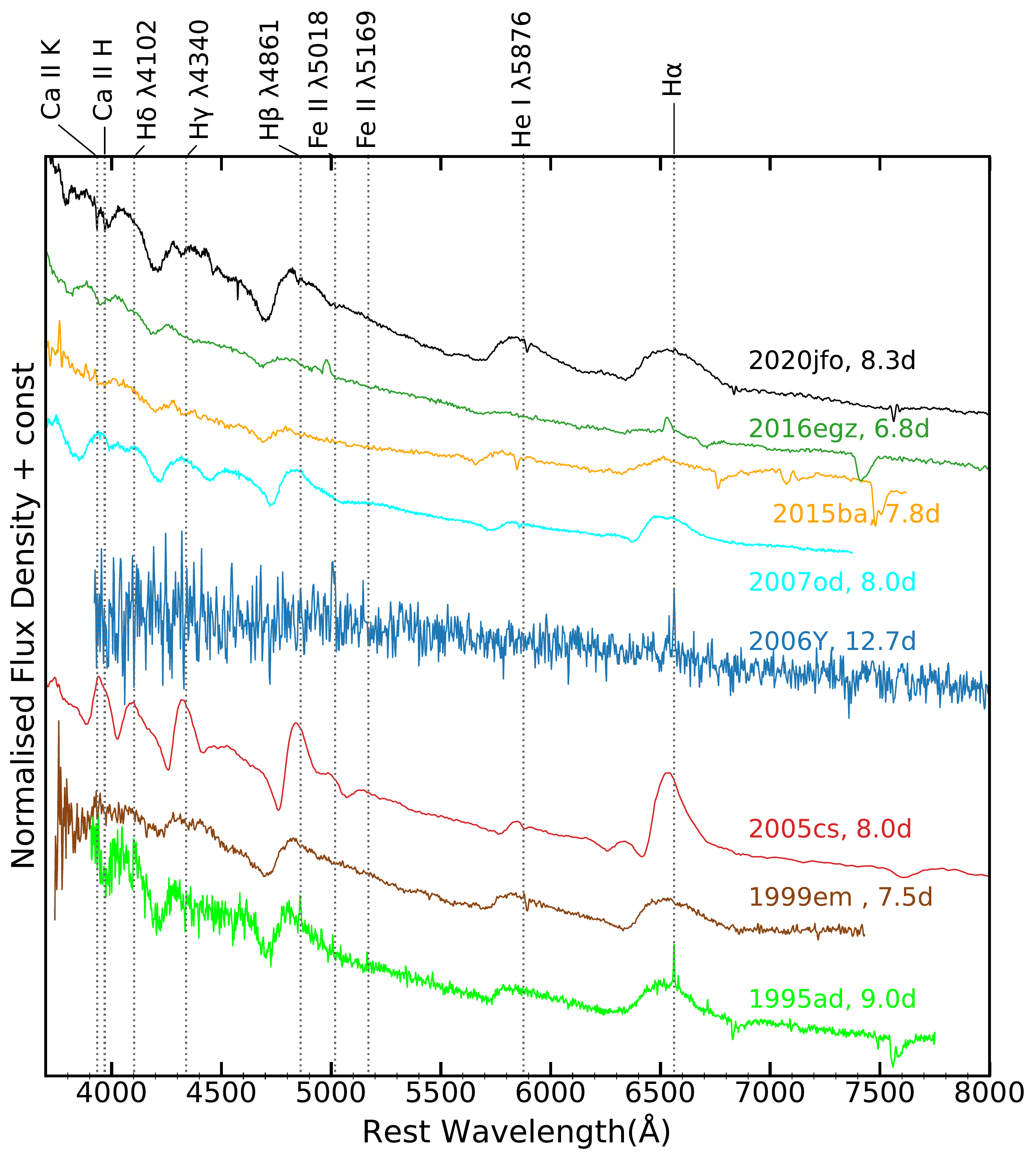}
    \caption{The spectrum of SN~2020jfo at $\sim 8$ day is compared with early-time spectra of other SNe~II in the comparison sample.}
    \label{early_spec_comp}
\end{figure}

At the early phase ($\sim 8$ day), Figure~\ref{early_spec_comp} indicates that the spectrum of SN~2020jfo displays broad H Balmer emission on a blue continuum similar to SNe~1995ad, 1999em, and 2007od. The comparable-epoch spectra of Type II SNe 2006Y (12.7 day), 2016egz (6.8 day), and SN 2015ba (7.8 day) are mostly blue and featureless corresponding to the shock-cooling phase. In contrast, the day 8 spectrum of SN~2005cs shows well-developed P~Cygni profiles of H Balmer lines, Ca~II H\&K, and feebly visible Fe~II, suggesting cooler photospheric temperatures for SN~2005cs.

\begin{figure*}
	\includegraphics[width=\textwidth]{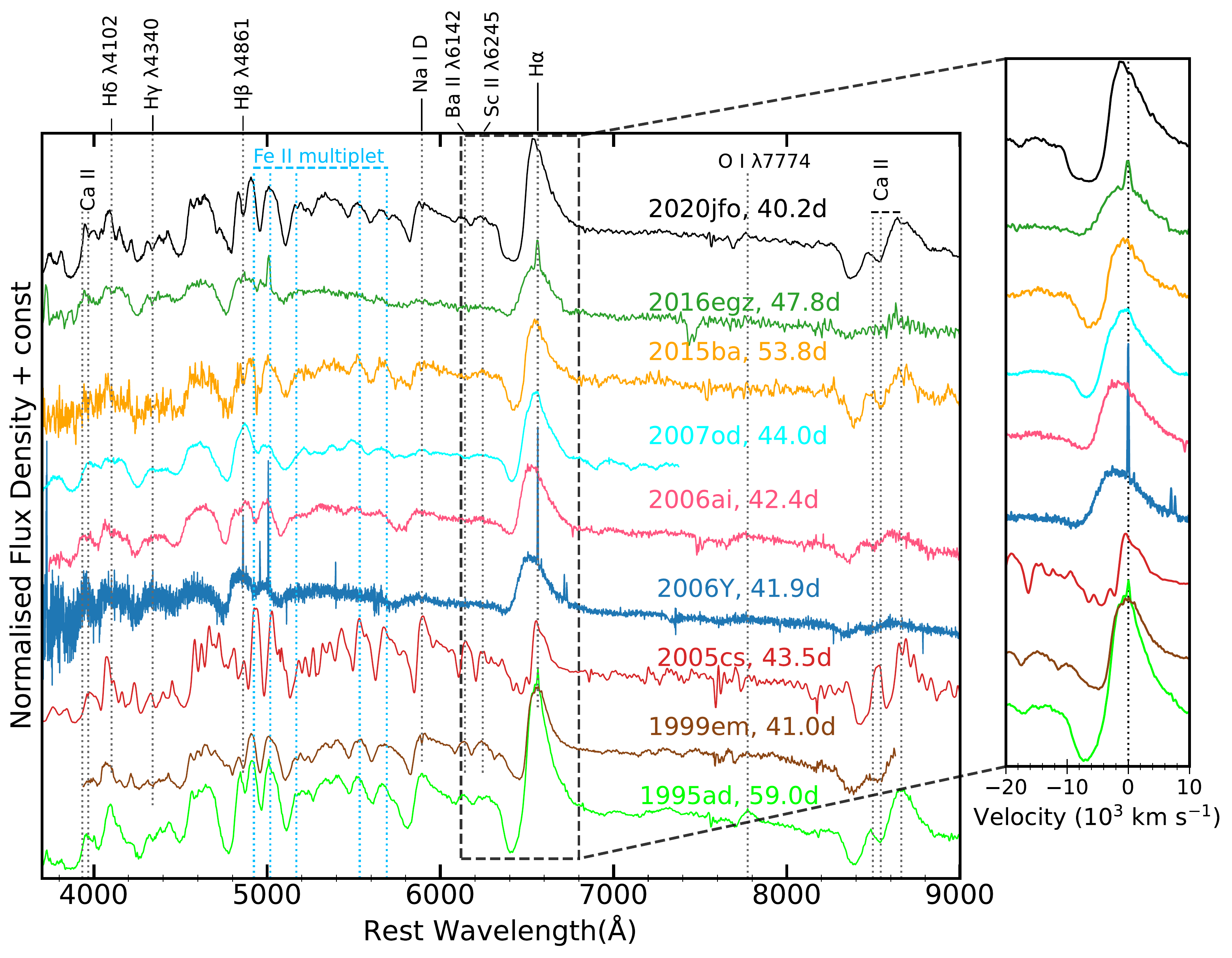}
    \caption{The plateau-phase spectrum of SN~2020jfo at $\sim 40$ day is compared with those of other SNe in the comparison sample. The wavelength region 6000--6800\,\AA\ is zoomed in and plotted in velocity space with respect to H$\alpha$, which highlights the similarity of the absorption component of H$\alpha$ in SN~2020jfo with the prototypical Type II SN 1999em.}
    \label{40_50}
\end{figure*}

In Figure~\ref{40_50}, we compare the plateau-phase spectrum ($\sim 40.2$ day) of SN~2020jfo with those of the comparison sample. The plateau-phase spectrum of SN~2020jfo is likewise similar to that of SNe~1995ad, 1999em, 2007od, and 2015ba. Unlike the other Type II SNe~2006Y and 2016egz, the spectrum of SN~2020jfo exhibits a substantial absorption component of the P~Cygni profile of H$\alpha$, which is shown by zooming in on the wavelength region 6000--6800\,\AA. The plateau spectrum of SN~2020jfo also shows well-developed Fe~II and feebly visible Sc~II and Ba~II lines. The Na I~D absorption is prominent in SN~2020jfo, similar to some Type II SNe with short plateau duration such as SNe~1995ad and 2006ai, but unlike the other short-plateau Type II SNe 2006Y, 2007od, and 2016egz. The Ca~II NIR triplet is also prominent in SN~2020jfo, similar to SNe~1995ad, 1999em, 2005cs, and 2015ba, and can be seen developing in the spectra of SNe~2006Y, 2006ai, and 2016egz.

\begin{figure*}
\includegraphics[width=\textwidth]{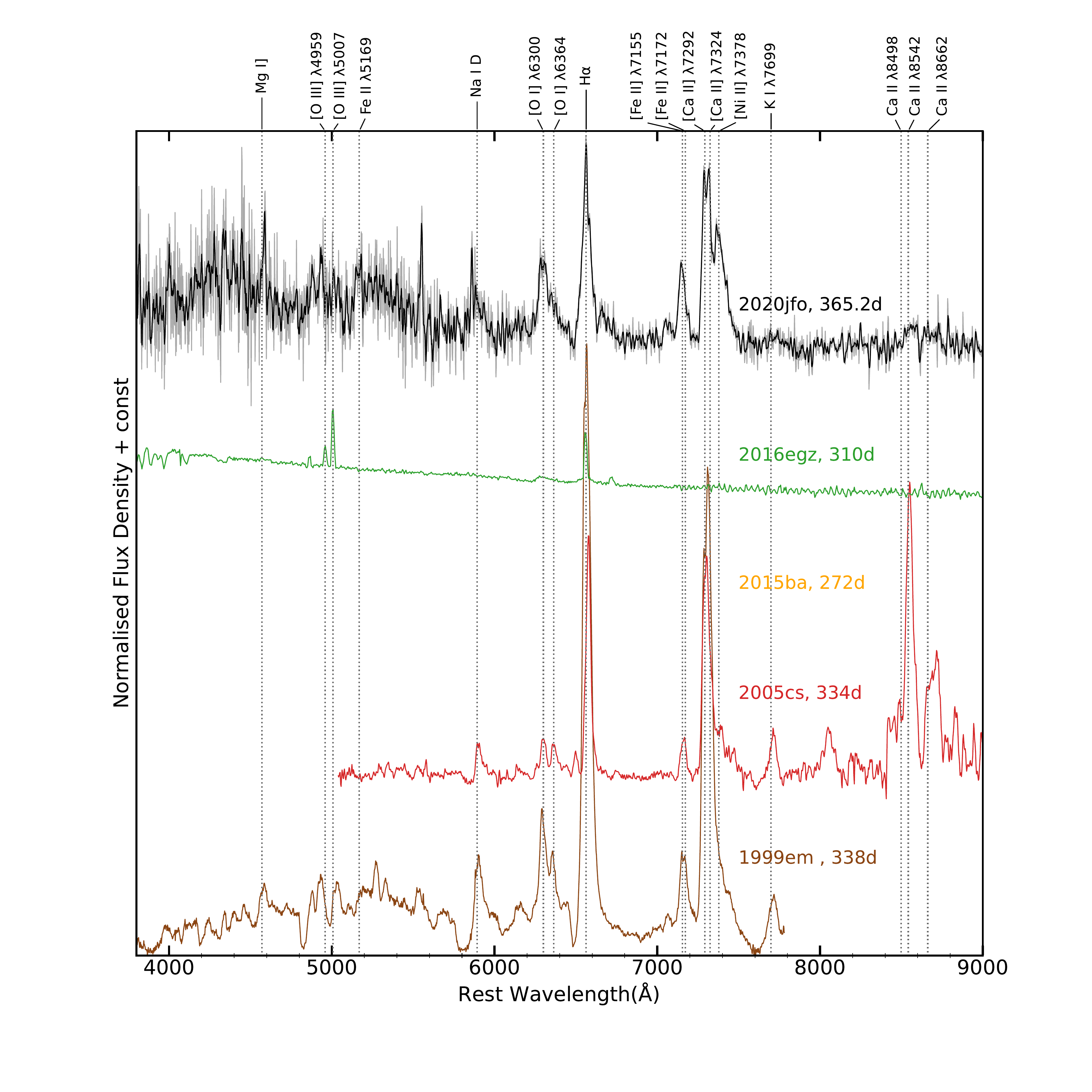}
    \caption{The nebular-phase spectrum of SN~2020jfo at $\sim 365$ day is compared with other SNe in the comparison sample. The nebular features in SN~2020jfo are similar to those seen in SN~1999em.}
    \label{late}
\end{figure*}

In Figure~\ref{late}, we compare the nebular ($\sim 365.2$ day) spectrum of SN~2020jfo with that of other SNe in the comparison sample at similar phases. Only one short-plateau SN 2016egz has a nebular spectrum after 300 days, which is mostly featureless. This is most likely due to host-galaxy contamination given its proximity to the host, as is also suggested by the the prominent narrow [O~III] $\lambda\lambda$4959, 5007 emission lines. The H$\alpha$ feature is pronounced in emission in all of the comparison spectra. The [O~I] $\lambda\lambda$6300, 6364 doublet and [Ca~II] $\lambda\lambda$7292, 7324 are very prominent in SN~2020jfo, similar to the other Type II SNe 1999em, 2005cs, and 2015ba. The nebular spectrum of SN~2020jfo exhibits strong [Ni~II] $\lambda$7378 emission which is indiscernible in the comparison sample. Overall, the features in the SN~2020jfo nebular spectrum are more similar to the spectra of SNe 1999em, 2005cs, and 2015ba, except the inconspicuous Ca~II NIR triplet in the spectrum of SN~2020jfo (unlike these other SNe). 

Thus, despite a short plateau duration, the spectral features of SN~2020jfo are more similar to those of normal Type II SNe. The length of the plateau phase is determined by the mass of the H envelope as well as the time it takes for the H to recombine. The progenitor of SN~2020jfo is likely to have an H envelope mass comparable to that of a typical SN~II, but with a shorter recombination time. However, rapidly expanding ejecta could explain the shorter recombination time; except for the first two epochs, the ejecta velocity of SN~2020jfo follows the mean velocity curve of normal Type II SNe (see Figure \ref{velocity trend}). The plateau duration, however, is also affected by the extent of $^{56}$Ni mixing in the ejecta, with a higher degree of mixing resulting in shorter plateaus \citep{Pumo2013, goldberg}.

\subsection{Velocity evolution}

\begin{figure}
	\includegraphics[scale=0.28]{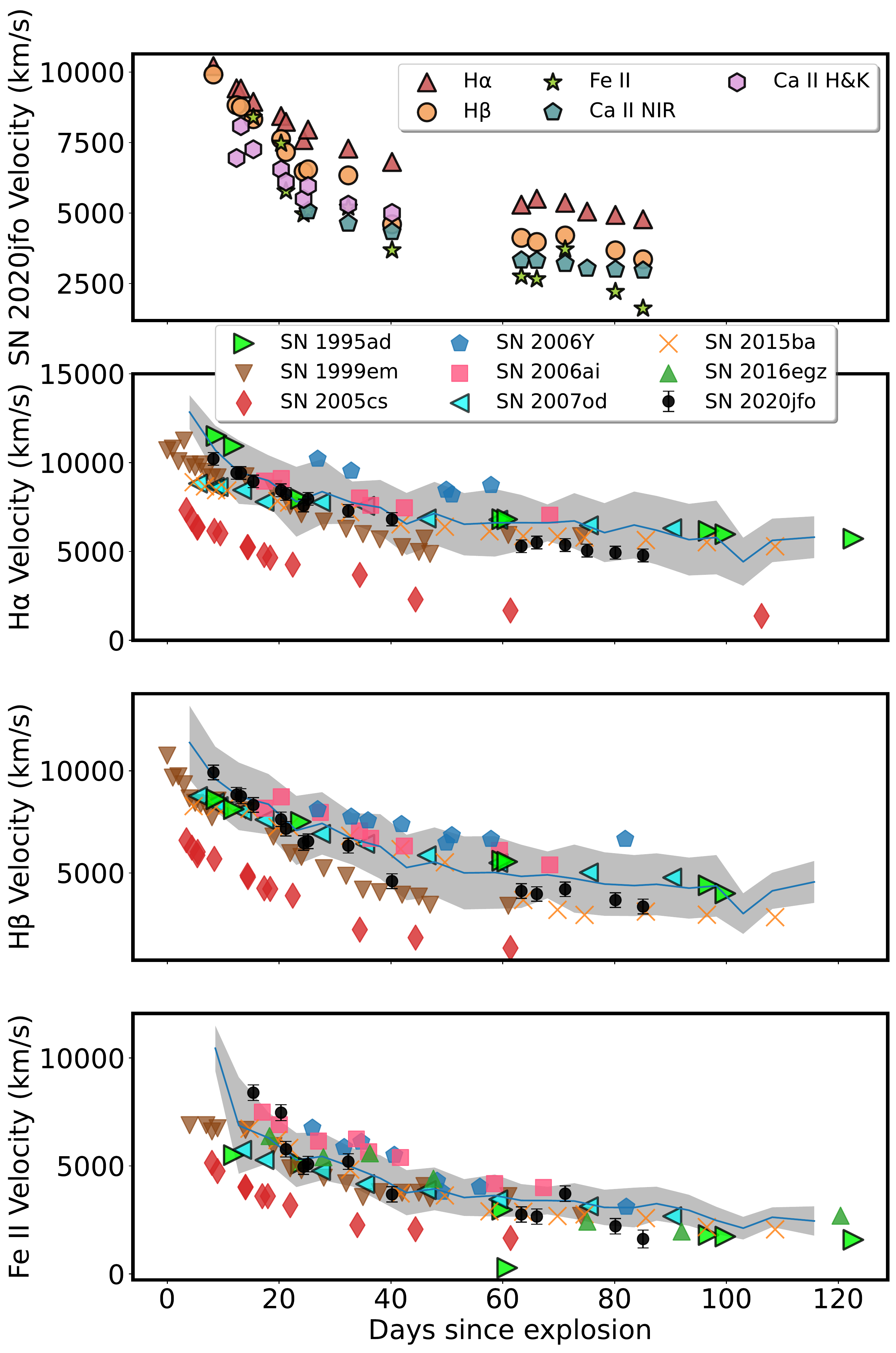}
    \caption{The evolution of line velocities of a few prominent lines in SN~2020jfo is shown in the top panel. The bottom three panels display a comparison of the velocity evolution of H$\alpha$, H$\beta$, and Fe~II lines of SN~2020jfo and the comparison sample. The mean velocities and the standard deviations of the velocities in the SN~II sample from \citet{gutirrez} are shown with a blue solid line and a grey band, respectively.}
    \label{velocity trend}
\end{figure}

The expansion velocity of the ejecta can be estimated from the minima of the absorption component of the P~Cygni profile of various lines. We estimate the line velocities of a few prominent features in SN~2020jfo and show their evolution with time in Figure \ref{velocity trend} (top panel). The  H$\alpha$ and H$\beta$ velocities are comparable at early phases (10,200 km s$^{-1}$ and 9900 km s$^{-1}$ at day 8) and greater than the velocity of other metal lines. As time proceeds, the H$\alpha$ velocity is higher than all other line velocities, showing it to be part of the outer expanding layer of ejecta. The velocity of the Fe~II line gives a fair estimate of the expansion velocity of the photosphere, and it is estimated to be around 8400 km s$^{-1}$ at $\sim 15$ days; it decays down to 1600 km s$^{-1}$ at the onset of the nebular phase. The decreasing trend of velocity shows that as time proceeds we are looking at the inner, lower-velocity layers of the ejecta owing to a decrease in the opacity resulting from the expansion of the ejecta.

The line velocities of SN~2020jfo are compared with those of other SNe in the sample. Figure~\ref{velocity trend} (bottom three panels) show the evolution of H$\alpha$, H$\beta$, and Fe~II line velocities. The mean velocity evolution and the standard deviation for a sample of 122 Type II SNe from \citet{gutirrez} is shown with a blue solid line and grey shaded region, respectively. Overall, the velocities of SN~2020jfo are lower than the velocities of the SNe in the comparison sample except SN~2005cs, and lie toward the edge of the grey band. The velocities of other SNe~II having shorter plateaus are typically higher than that of SN~2020jfo.  The Fe~II velocity of SN~2020jfo at day 15 after explosion is $\sim 1900$ km s$^{-1}$ higher than the mean velocity.

\section{Progenitor and explosion properties}
\label{progenitor}

In this section, we adopt different techniques to estimate the progenitor properties, and we provide a cohesive picture of this analysis in Section \ref{conclusions}.

\subsection{Estimations from the nebular spectrum}
\label{sec:prog_spec}

\begin{figure*}
	\includegraphics[width=\textwidth]{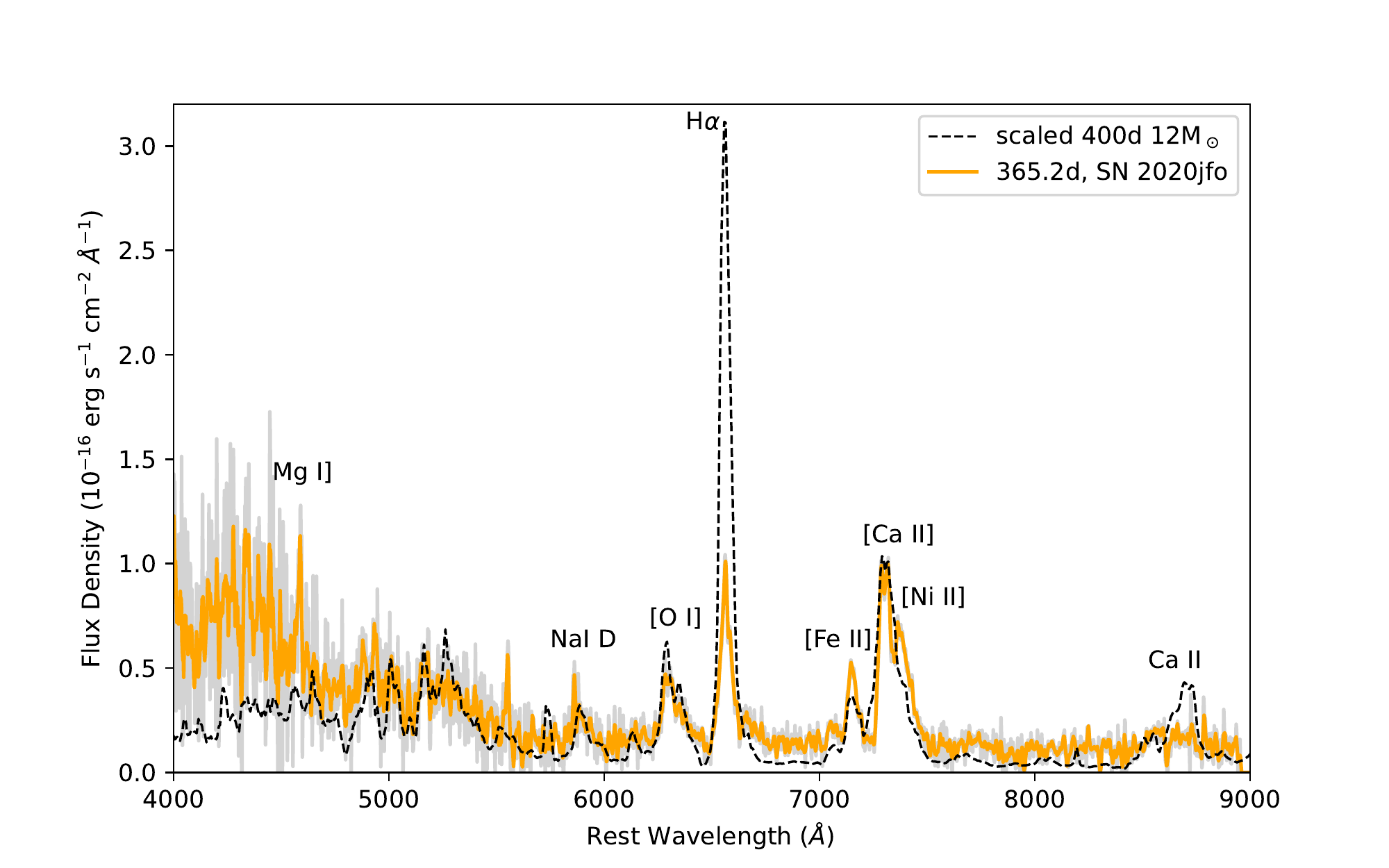}
    \caption{Comparison of the day 365.2 spectrum of SN~2020jfo with that of the scaled model spectra from \citep{jerkstrand}.}
    \label{fig:Jerkstrand_comp}
\end{figure*}

The observed strengths of nebular lines, especially the [O~I] $\lambda\lambda$6300, 6364 doublet, can serve as a diagnostic of the progenitor mass, as the nucleosynthetic yield depends strongly on the mass of the progenitor \citep{woosley}. \cite{jerkstrand} generated models of nebular spectra for 12, 15, 19, and 25 M$_\odot$ progenitors for 0.062 M$_\odot$ of $^{56}$Ni, assuming a distance of 5.5 Mpc. We compare the day 365.2 spectrum of SN~2020jfo to these models to get a handle on the mass of the progenitor as shown in Figure~\ref{fig:Jerkstrand_comp}. As the SN is powered by the radioactive decay energy in the nebular phase, we scaled the day 400 model spectrum flux with the ratio of the $^{56}$Ni mass of SN 2020jfo (0.030 M$_\odot$) to the $^{56}$Ni mass of the model spectrum (0.062 M$_\odot$). 
Then the day 400 model flux was rescaled to the distance and phase (365.2 day) of the nebular spectrum of SN~2020jfo. We also used a scaling factor of 0.5, similar to \cite{sollerman}, owing to the partial trapping of $\gamma$-rays in SN~2020jfo. Before the comparison, we deredshifted the SN spectrum, scaled the spectrum to match the photometric flux so that the slit losses are taken care of, and applied reddening corrections. The SN spectrum best matches with the 12 M$_\odot$ model. However, the H$\alpha$ flux in the spectrum of SN~2020jfo is a factor of 3 lower than the model flux. The model also shows a prominent Ca~II NIR triplet feature which is absent in the SN~2020jfo spectrum. The [O~I] flux in the SN~2020jfo spectrum is comparable to the model spectrum, suggesting a progenitor ZAMS mass of $\sim 12$ M$_\odot$ for SN~2020jfo. We note here that \cite{sollerman} and \cite{teja} also carried out this comparison with the nebular spectra of SN 2020jfo and our result is consistent with theirs.

The relative abundances of the different elements have been suggested to vary substantially with the helium core mass \citep{fransson1989}; specifically, the [O~I]/[Ca~II] ratio has been used to estimate the progenitor's ZAMS mass by several  authors (e.g., \citealt{fransson1989}; \citealt{elmhamdi2004}; \citealt{Kuncarayakti}). We estimated the [O~I] and [Ca~II] luminosity in the nebular spectrum (365.2 day) of SN~2020jfo to be $(1.13 \pm 0.19) \times 10^{38}$ erg s$^{-1}$ and $(1.96 \pm 0.06) \times 10^{38}$ erg s$^{-1}$, respectively, and the [O~I]/[Ca~II] ratio is $0.58 \pm 0.10$. This ratio has been found to fall below 0.7 for Type II SNe, while for other CCSNe, the ratio is higher than 0.7 \citep{Kuncarayakti}. Thus, for SN~2020jfo, the [O~I]/[Ca~II] ratio coincides with the upper limit for SNe~II. The fact that the ratio increases with increasing main-sequence mass suggests that the progenitor of SN~2020jfo falls on the upper end of the mass distribution for Type IIP SN progenitors.

\subsection{Ni/Fe production ratio}

Ni and Fe are essentially produced in the same region of the ejecta in Fe CCSNe, and the [Fe~II] emission is stronger compared with [Ni~II] in these events \citep{Jerkstrand2012}. \citet{Jerkstrand2015J} presented an analytic method to determine the Ni/Fe ratio in nebular spectra of SNe in the 7100--7400\,\AA\ region, which we have applied here to SN~2020jfo. There are eight prominent emission lines in the 7100--7500\,\AA\ region: [Ca~II] $\lambda\lambda$7291, 7323, [Fe~II] $\lambda$7155, [Fe~II] $\lambda$7172, [Fe~II] $\lambda$7388, [Fe~II] $\lambda$7453, [Ni~II] $\lambda$7378, and [Ni~II] $\lambda$7412. We fit this spectral region in the day 253.9, 255.9, and 365.2 spectra of SN~2020jfo using Gaussian components for these lines. Following the Jerkstrand model, we fixed the line ratios from the same species. Thus, the iron lines are constrained by $L_{7453} = 0.31\,L_{7155}$, $L_{7172} = 0.24\,L_{7155}$, and $L_{7388} = 0.19\,L_{7155}$, and the nickel lines are constrained by $L_{7412} = 0.31\, L_{7378}$. Furthermore, in this method, the full width at half-maximum  (FWHM) velocity $v$ of all the emission lines are forced to be the same for a given epoch, so that there are four free parameters: $L_{7291, 7323}$, $L_{7155}$, $L_{7378}$, and $v$. 

\begin{figure}
	\includegraphics[width=\columnwidth]{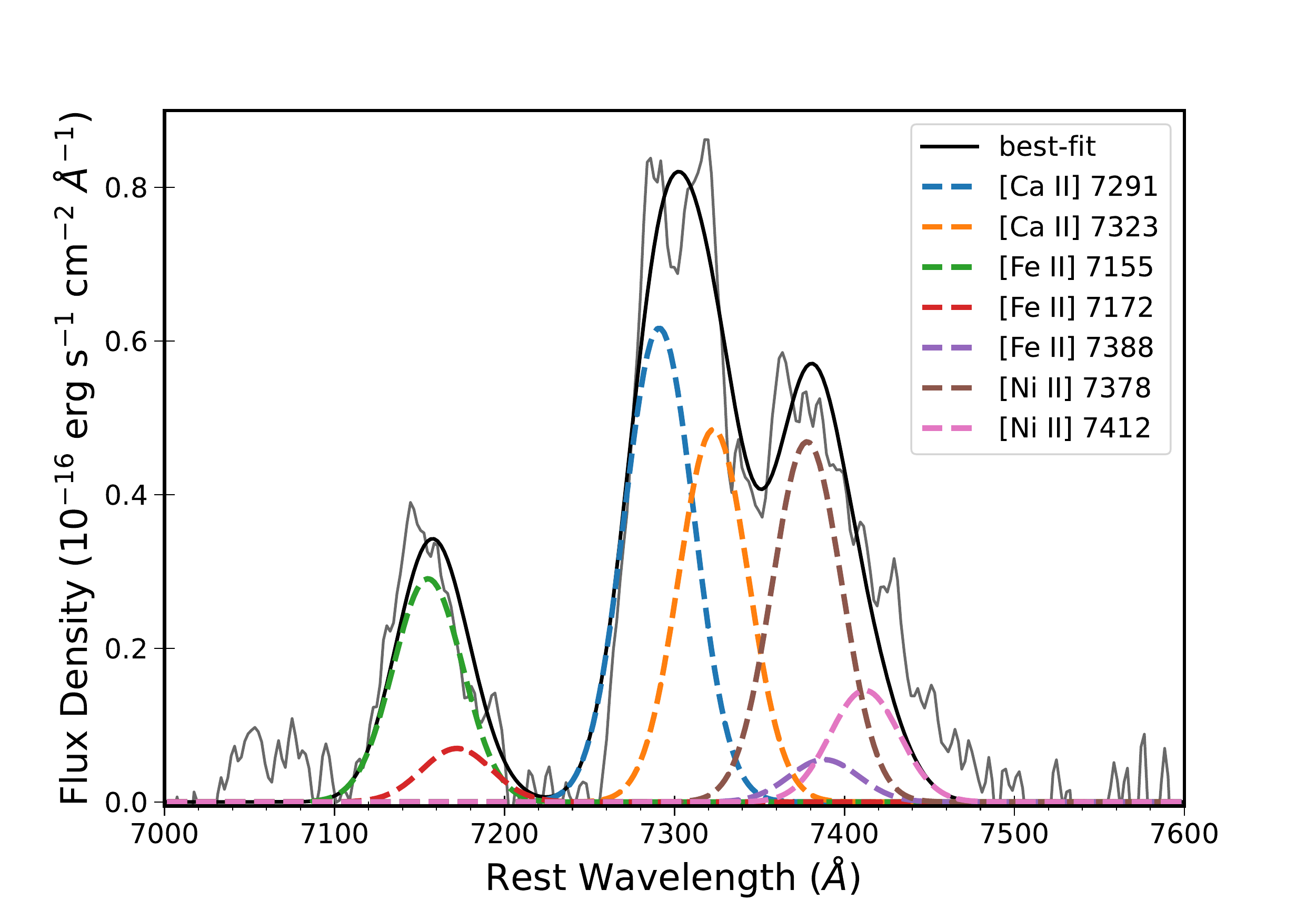}
    \caption{Section of the 365.2 day spectrum showing the Gaussian fits to determine [Ni~II] $\lambda$7378/[Fe~II] $\lambda$7155 in SN~2020jfo.}
    \label{fig:Ni_Fe_ratio}
\end{figure}

In Figure~\ref{fig:Ni_Fe_ratio}, we show that a good fit in the day 365.2 spectrum is obtained for $L_{7291, 7323} = (1.96 \pm 0.06) \times 10^{38}$ erg~s$^{-1}$, $L_{7155} = (0.57 \pm 0.02) \times 10^{38}$ erg~s$^{-1}$, $L_{7378} = (0.84 \pm 0.03) \times 10^{38}$ erg~s$^{-1}$, and $v = 2158$ km~s$^{-1}$. From this, we determine a ratio $L_{7378} /L_{7155} = 1.61 \pm 0.11$ on day 365.2. We followed the same approach and estimated the luminosity ratio of [Ni~II] to [Fe~II] to be $1.65 \pm 0.18$ and $1.98 \pm 0.21$ in the day 253.9 and 255.9 spectra of SN~2020jfo, respectively. The luminosity ratio measured by \cite{sollerman} with a day 306 spectrum is 1.7 and by \cite{teja} with a day 292 spectrum is $2.10 \pm 0.43$. Using $T = 2700$\,K, we obtain a Ni-to-Fe mass ratio in the range 0.08--0.10, which is $\sim 1.5$ times the solar value. We see that [Ni~II] $\lambda$7378 emission is stronger than [Fe~II] $\lambda$7155 in SN~2020jfo, which is usually seen in low-luminosity SNe~II \citep{Muller-Bravo2020}. While we derived a mean ratio using three nebular spectra, both \cite{sollerman} and \cite{teja} used a single nebular spectrum and obtained a higher value of this mass ratio. The ratio estimated using the day 306 spectrum by \cite{sollerman} is 2 times the solar value and that by \cite{teja} using the day 292 spectrum is 3 times the solar value; however, in the latter work, the temperature used to estimate this quantity is not mentioned.

\subsection{Bolometric light curve modelling}

\begin{figure}
	\includegraphics[scale=0.55]{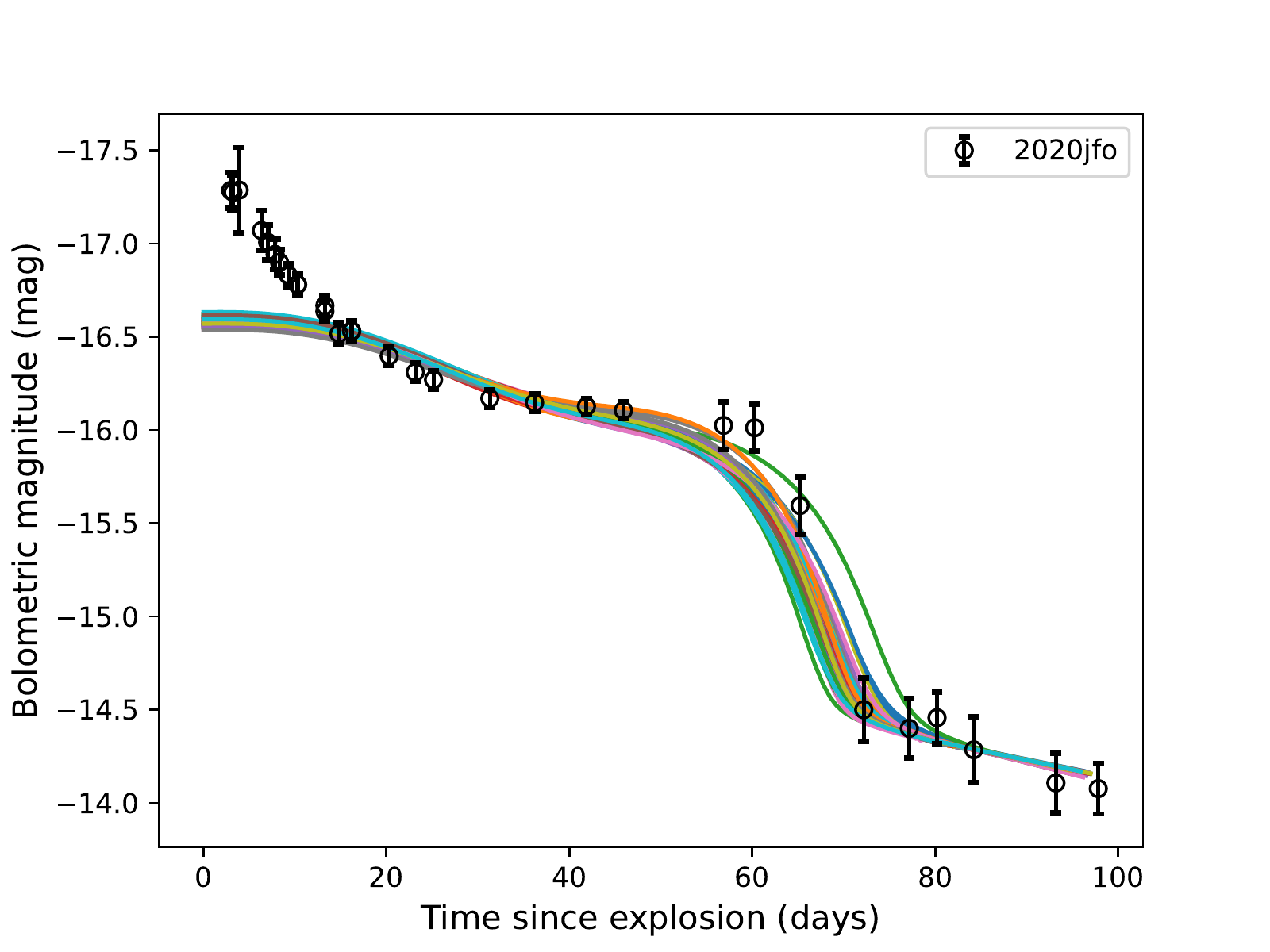}
    \caption{The bolometric magnitude evolution of SN~2020jfo is shown with the 50 best-fit light curves overplotted, following \citet{2016A&A...589A..53N} and \citet{2020MNRAS.496.3725J}.}
    \label{fig:Nagy}
\end{figure}

We constructed the bolometric light curve of SN\,2020jfo using {\sc superbol} \citep{2018RNAAS...2..230N} as shown in Figure\,\ref{fig:Nagy}. The dereddened magnitudes supplied to the routine are interpolated to a common set of epochs and converted to fluxes. The fluxes are used to construct the spectral energy distribution (SED) at all the epochs. The routine then fits a blackbody function to the SED, extrapolating to the UV and IR regimes to estimate the true bolometric luminosity. For SN\,2020jfo, we used the \textit{Swift} $uvw2$, $uvm2$, and $uvw1$ bands, optical $BgVri$ bands, and IR $J$-band data from \cite{sollerman} to construct the first 50 days bolometric light curve using {\sc superbol}. For the rest of the light curve, we used only $BgVri$ data. 

We employed the semi-analytical modelling of \cite{2016A&A...589A..53N} to which an MCMC method was added by \cite{2020MNRAS.496.3725J} for numerically fitting the output model to the observed light curves. \cite{2016A&A...589A..53N} used a core+shell combination to fit the early declining part with the shell configuration. The plateau phase, drop, and tail are modelled with the core configuration. However, the MCMC method added by \cite{2020MNRAS.496.3725J} only takes into account the core part, and hence the early declining part (the first 10 days in the light curve) is not fitted using this method. For the modelling, we have sampled five parameters: radius ($R_0$), ejecta mass ($M_{\mathrm{ej}}$), kinetic energy ($E_{\mathrm{kin}}$), thermal energy ($E_{\mathrm{th}}$), and opacity. A mean and sigma prior expansion velocity of 8140 km~s$^{-1}$ and 500 km~s$^{-1}$ (respectively) were used, along with a recombination temperature of 5500~K. From the model, we estimated an initial radius ($R_0$) of $\sim 1000$ R$_\odot$, ejecta mass ($M_{\mathrm{ej}}$) of 13.6 M$_\odot$, and total energy ($E_{\mathrm{kin}}+E_{\mathrm{th}}$) of 3.0 foe. The bolometric magnitude evolution and the 50 best-fit light curves are shown in Figure~\ref{fig:Nagy}, and the corresponding parameter estimates and their 1$\sigma$ uncertainties are provided in Table \ref{Nagy}. 

There are two known parameter correlations \citep{1989ApJ...340..396A, 2014A&A...571A..77N} between $M_{\mathrm{ej}}$ and $E_{\mathrm{kin}}$, and between $R_0$ and $E_{\mathrm{th}}$. Since the two parameter pairs $M_{\mathrm{ej}} - E_{\mathrm{kin}}$ and $E_{\mathrm{th}} - R_0$ are significantly correlated, separate determination of the quantities in the pair is not possible. Rather, the product of the pairs should be used as an independent parameter. The $^{56}$Ni mass estimated from the modelling using the tail light curve is 0.022\,M$_\odot$. Assuming a remnant neutron star mass of 1.5--2\,M$_\odot$, the lower limit of the progenitor mass would be $\sim 15$\,M$_\odot$, which lies within the expected mass range of RSGs exploding as Type IIP SNe. We note here that we could not reproduce the early luminosity excess by solely using the core component in the semi-analytical model. Thus, we cannot rule out the possibility of CSM interaction from pre-explosion mass loss.

In \citet{Inserra1995ad}, hydrodynamic modelling of the bolometric light curve of SN~1995ad was performed, which estimated 5\,M$_\odot$ of ejecta mass and a progenitor radius of $4 \times 10^{13}$ cm. They proposed both super asymptotic giant branch (SAGB) and Fe~CC progenitors, and noted that the former is less likely owing to the relatively high amount of $^{56}$Ni ejected. The range of ejecta mass (5--7.5\,M$_\odot$) and progenitor radius ((4--7) $\times 10^{13}$ cm) estimated for SN~2007od are also similar to SN~1995ad. In the work of \citet{hiramatsu}, three short-plateau SNe 2006ai, 2006Y, and 2016egz were studied and the ZAMS mass ($M_{\rm ZAMS}$) of the progenitor was suggested to be in the range $\sim 18$--22\,M$_\odot$, much higher than that of SNe~1995ad and 2007od. The study indicated that the progenitors of these SNe~II had experienced massive stripping prior to explosion \citep{hiramatsu}, leading to a very low H-envelope mass ($\sim 1.7$\,M$_\odot$), much lower than that of SNe~1995ad and 2005od. The ejecta mass estimated from semi-analytical modelling, which mainly constitutes the H-envelope mass, for SN~2020jfo is much higher than that of the SNe~II studied by \citet{Inserra1995ad}, \citet{inserra2011}, and \citet{hiramatsu}. Although the early luminosity excess in SN~2020jfo and the high-velocity feature in H$\alpha$ indicate pre-explosion mass loss, the envelope stripping in the progenitor of SN~2020jfo is expectedly much less violent than in the Type II SNe 2006Y, 2006ai, and 2016egz.

\begin{table}
\centering
\caption{The best-fit core parameters and 1$\sigma$ confidence intervals for the bolometric light curve of SN\,2020jfo using \citet{2016A&A...589A..53N}.
\label{Nagy}}
\setlength{\tabcolsep}{10pt} 
\renewcommand{\arraystretch}{1.5} 
\begin{tabular}{lcl}
\hline
$R_0$ ($10^{13}$~cm)       & $6.89^{+0.45}_{-2.78}$        & Initial radius of ejecta\\
$M_{\mathrm{ej}}$ (M$_\odot$)       &  $13.6^{+0.2}_{-2.5}$     & Ejecta mass\\
$E_{\mathrm{kin}}$ (foe)            & $2.9^{+0.4}_{-0.5}$   & Initial kinetic energy \\
$E_{\mathrm{th}}$ (foe)             & $0.066^{+0.038}_{-0.004}$     & Initial thermal energy \\
\hline
\end{tabular}
\end{table}

\subsection{Metallicity}
\label{sec:metallicity}

Metallicity is one of the key parameters that constrains the nature of the stellar evolution and the mechanism of the SN explosion, particularly in the case of high-mass stars (e.g., \citealt{heger}). \cite{pilyugin} derived an empirical relation between absolute $B$-magnitude ($M_B$) and metallicity (12 + log($N_O/N_H$)) for the nearby spiral galaxies (along with M61, the host of SN~2020jfo) and computed the radial profile of their metallicity distribution. The nuclear metallicity of M61 is 8.84 dex, and the angular separation between the SN and the host center is $\sim 1.2'$. This implies that the metallicity at the SN location is roughly 8.57 dex (8.65 dex is the solar-neighbourhood metallicity; \citealt{asplund}). Thus, the estimated metallicity of the region of SN~2020jfo using this method appears to be nearly solar, although it is higher than that of SN~2008in (8.44 dex; \citealt{roy}) which was located in the same host galaxy.

\begin{figure}
	\includegraphics[width=\columnwidth, clip, trim ={0cm 2cm 0cm 3cm}]{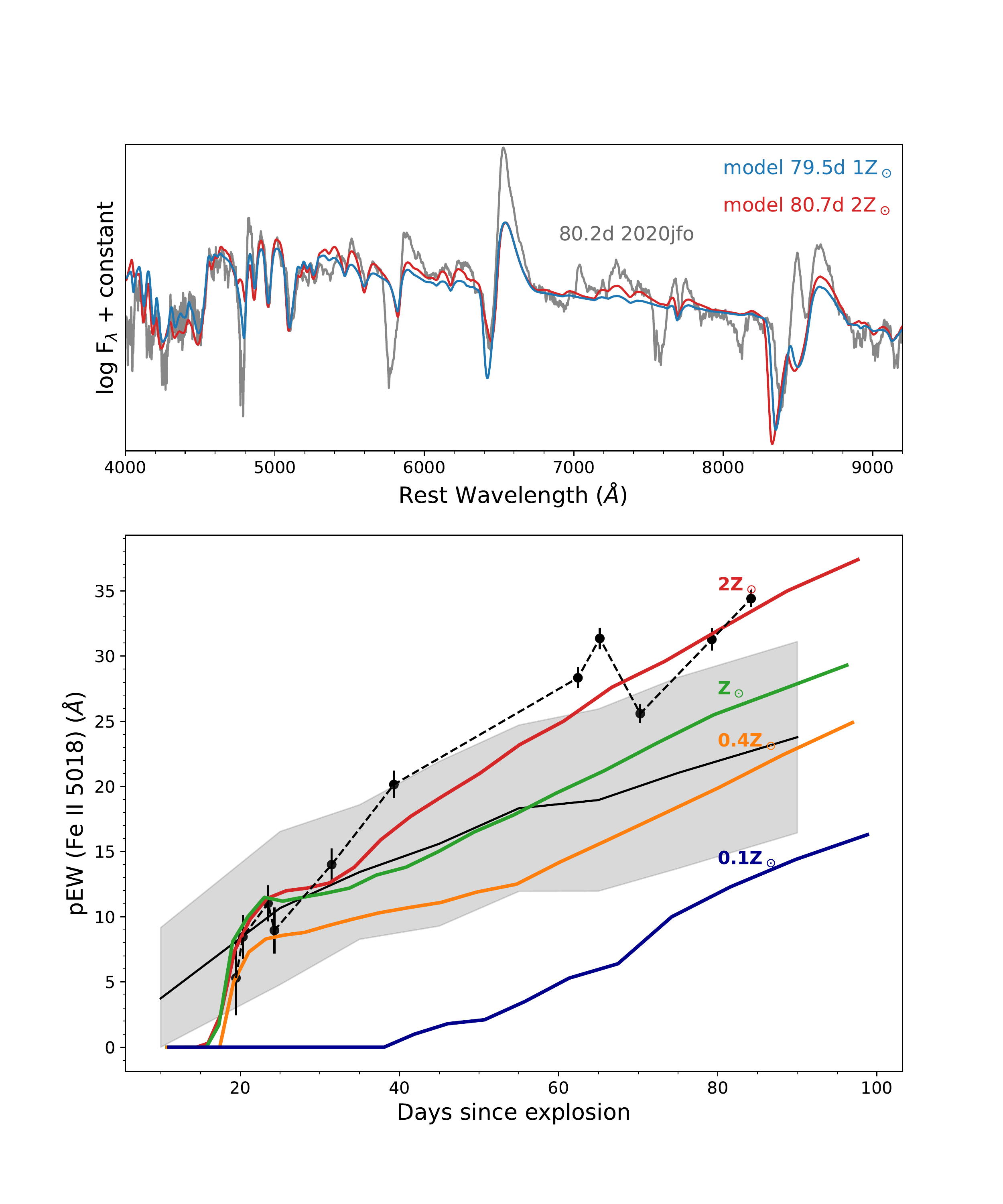}
    \caption{{\bf Top panel:} Comparison of the day 80.2 spectrum of SN~2020jfo with the best-matched model spectrum of \citet{dessart2013}. {\bf Bottom panel:} The evolution of the pEW of Fe~II $\lambda$5018 of SN~2020jfo is shown with dashed lines. The black line is the mean pEW of a sample of 119 Type II SNe from \citep{anderson} and the shaded region shows the standard deviation about the mean. The time evolution of the pEWs of Fe~II $\lambda$5018 of the four distinct metallicity models of \citep{dessart2013} is shown with thick lines.}
    \label{metallicity}
\end{figure}

Metallicity also plays a crucial role in the evolution of metal lines in the photospheric phase as the SNe with low progenitor metallicity exhibit weaker metal features \citep{dessart2014}. \cite{dessart2013} simulated model spectra by evolving a 15\,M$_\odot$ ZAMS star up to the pre-SN stage at various metallicities (0.1, 0.4, 1.0, 2.0\,Z$_\odot$). We compared the day 80.2 spectrum of SN~2020jfo with these models and found a reasonable match with both 1 and 2\,Z$_\odot$ model spectra considering the Fe~II $\lambda$5018 and $\lambda$5169 features, that are overplotted on the spectrum of SN~2020jfo in the top panel of Figure~\ref{metallicity}. However, the higher metallicity model matches the H$\alpha$ absorption better. Furthermore, \cite{anderson} estimated the oxygen abundances using a sample of 119 host H~II regions and found a positive correlation between the host H~II region oxygen abundance and the pseudo-equivalent width (pEW) of Fe~II $\lambda$5018. In the bottom panel of Figure~\ref{metallicity}, the time evolution of the mean pEW of Fe~II $\lambda$5018 of the sample of Type II SNe from \cite{anderson} is shown along with its standard deviation (the black line and the shaded region), over which the thick solid lines are the time evolution of the pEW of Fe~II $\lambda$5018 in the different metallicity models of \cite{dessart2013}. The pEW measurements of Fe~II $\lambda$5018 in the spectra of SN~2020jfo is shown with a dashed line which falls slightly above the evolution of the 2\,Z$_\odot$ model and is higher than the mean pEW of the sample. This estimate is higher than that obtained from radially decreasing metallicity gradients in spiral galaxies. Generally, higher metallicity would imply a higher extent of envelope stripping of the progenitor prior to explosion \citealt{heger}. Despite the limitations of both methods of metallicity estimation, the higher metallicity scenario is more consistent with SNe~II having shorter plateaus, where the progenitors are expected to have undergone comparatively extensive envelope stripping than Type II SNe with long plateaus \citep{hiramatsu}. In the context of SN~2020jfo, however, our analysis does not indicate significant envelope stripping. Thus, we prefer the low-luminosity scenario for SN~2020jfo.

\section{Collocation of SN~2020jfo in the SN II parameter space}
\label{collocation}

We explored the collocation of SN~2020jfo in the parameter space of other SNe~II for which we have used the sample of Type II SNe from \cite{anderson14}, \cite{faran14}, and \cite{valenti16}. 

\begin{figure}
	\includegraphics[width=\columnwidth]{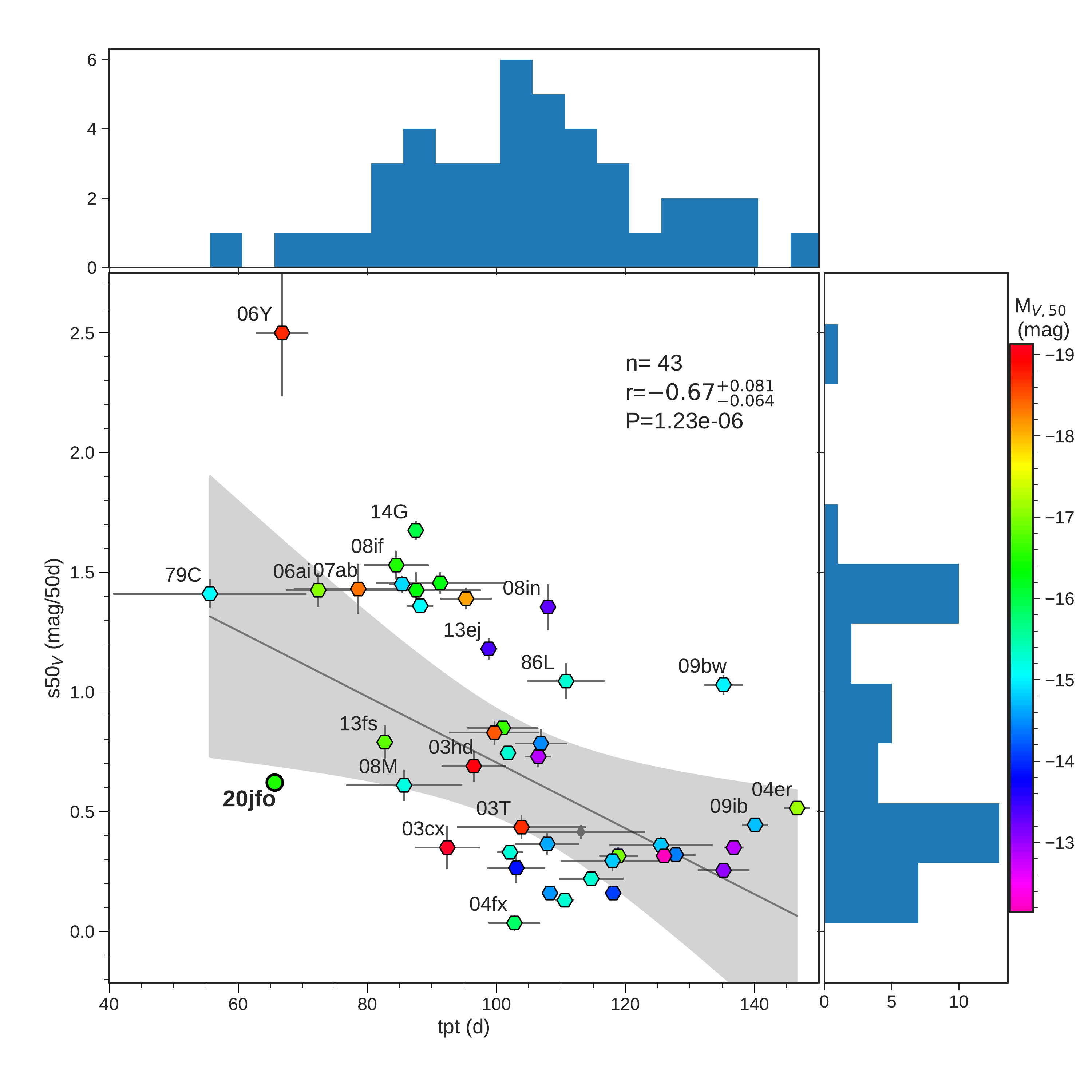}
    \caption{The correlation between $s50V$ and $t_{\rm pt}$ is shown. The objects are colour-coded with respect to the absolute $V$-band magnitude at day 50.}
    \label{fig:corr2}
\end{figure}

In Figure~\ref{fig:corr2}, the plateau decline rate in the $V$ band and $t_{\rm pt}$ of SN~2020jfo are shown colour-coded with the $V$ absolute magnitude at day 50, along with that of other SNe. There is a correlation among these parameters as noted by \cite{valenti16}. The correlation implies that SNe with longer plateaus tend to decline more slowly. The SNe with $t_{\rm pt}$ less than 80 days are SNe 1979C, 2006Y, and 2006ai. Although the plateau lengths of SNe 2006ai and 2006Y are consistent with SN~2020jfo within the uncertainties, SN 2006Y is brighter than SNe~2020jfo and 2006ai. We performed a linear fit and have shown the 3$\sigma$ confidence interval with the grey shaded region. SN 2006ai falls within the 3$\sigma$ confidence interval whereas SN~2020jfo falls outside it. Among the SNe with a short plateau duration, SN~2020jfo has the shallowest decline rate. In general, the SNe with shorter plateaus are brighter, with a higher decline rate than those with longer plateau. SNe with absolute plateau magnitudes in the range $-18$ to $-19$ mag are spread over all decline rates, while for the rest of the sample, SNe with lower luminosity preferably have shallower decline rates. SN~1979C is considered to be a prototypical Type IIL SN that does not show a clear light-curve drop. As suggested by \cite{anderson14}, SN~1979C may have a light-curve drop around day 50 after explosion, though if there, the drop would have occurred quite early and less pronounced than in all the other cases.

We plot the absolute magnitude in $V$ at day 50 against the logarithm of the $^{56}$Ni mass for a sample of Type II SNe. We find that the SNe~II  with higher luminosity (and hence higher explosion energy) have a larger $^{56}$Ni yield. SN~2020jfo follows the correlation between these two parameters. Also, we plot peak magnitude in $V$ against $a_0$ of SN~2020jfo along with those of other Type II SNe. In general, SNe with higher peak magnitudes have shorter plateau lengths. Hence, the fall in magnitude from the plateau to the radioactive tail phase is expected to be lower. SN~2020jfo follows the trend, but the plateau length of SN~2020jfo is shorter and slowly declining compared with other short-plateau SNe.

The correlations suggest that although SN~2020jfo has a short plateau length, its parameters are more similar to those of normal SNe~IIP.

\section{Conclusions}
\label{conclusions}

In this paper, we present a photometric and spectroscopic analysis of the Type II SN~2020jfo. Using the light-curve equation from \citet{valenti16}, we estimated a short-plateau length of $67.0 \pm 0.6$ days for SN~2020jfo, which is consistent with the result given by \cite{sollerman} and \cite{teja}. 
 
However, unlike other SNe~II with shorter plateau duration, SN~2020jfo is fainter, with a peak absolute $V$-band magnitude of $-16.90 \pm 0.34$ mag, in comparison with the luminous optical peaks of those SNe ($\lesssim -18.4$ mag). While the plateau decline rates in $V$ of SN~2020jfo and other SNe~II having shorter plateau duration are similar, the short-plateau Type II SNe 1995ad, 2006Y, 2006ai, 2007od, and 2016egz display a steeply declining phase in $V$ up to $\sim 20$ days, which is not observed in SN~2020jfo. The $^{56}$Ni mass yield of SN~2020jfo estimated from the tail-phase luminosity is $0.03 \pm 0.01$\,\(M_\odot\), which falls in the range of normal SNe~II. 

Despite the shorter plateau duration in SN~2020jfo, the H~I absorption features in its plateau-phase spectra are remarkably strong, indicating a relatively high H-envelope mass. SN~2020jfo also shows strong metal lines in comparison with other SNe~II at similar epochs and having comparable plateau lengths. Moreover, a discernible line of stable [Ni~II] $\lambda$7378, which has only been seen in a few H-rich SNe, could be identified in the nebular spectra of SN~2020jfo. Analysis of the line ratio of [Ni~II] $\lambda$7378 to [Fe~II] $\lambda$7155 suggests a supersolar Ni-to-Fe production ratio by mass.

From nebular-phase spectral comparison \citep{jerkstrand}, we estimated the ZAMS mass of the progenitor of SN~2020jfo to be around 12\,M$_\odot$. Moreover, semi-analytical modelling using the prescription of \cite{2016A&A...589A..53N} indicated an ejecta mass of $13.6^{+0.2}_{-2.5}$ M$_\odot$. Adding a remnant neutron star mass, the progenitor mass turns out to be $15^{+0.2}_{-2.5}$ M$_\odot$. This can be considered as a lower limit of the progenitor ZAMS mass, since the analytical method does not estimate the mass lost by winds. The SN~2020jfo ejecta mass estimated from semi-analytical modelling, which mainly constitutes the H-envelope mass, is much higher than that of the Type II SNe studied by \citet{Inserra1995ad}, \citet{inserra2011}, and \citet{hiramatsu} with similar plateau lengths. Thus, spectroscopic comparison and modelling suggests that SN~2020jfo differs from SNe~II with similar plateau lengths and is rather spectroscopically similar to SN~1999em. 

\section*{Acknowledgements}
The authors thank the referee for critically reviewing the manuscript which has improved the presentation of the paper. This research made use of the NASA/IPAC Extragalactic Database (NED) that is operated by the Jet Propulsion Laboratory, California Institute of Technology, under contract with the National Aeronautics and Space Administration (NASA). This work makes use of data obtained by the LCO network. Based in part on observations obtained at the 3.6\,m Devasthal Optical Telescope (DOT), which is a National Facility run and managed by Aryabhatta Research Institute of Observational Sciences (ARIES), an autonomous Institute under Department of Science and Technology, Government of India. We thank (mostly U.C. Berkeley undergraduate students) Raphael Baer-Way, Matthew Chu, Asia deGraw, Nachiket Girish, Evelyn Liu, Emily Ma, Shaunak Modak, Derek Perera, and Keto D. Zhang for their effort in taking Lick/Nickel data. The Kast red CCD detector upgrade, led by B.  Holden, was made possible by the Heising-Simons Foundation, William and Marina Kast, and the University of California Observatories. KAIT and its ongoing operation were made possible by donations from Sun Microsystems, Inc., the Hewlett-Packard Company, AutoScope Corporation, the Lick Observatory, the US National Science Foundation, the University of California, the Sylvia \& Jim Katzman Foundation, and the TABASGO Foundation. Research at Lick Observatory is partially supported by a generous gift from Google. 

B.A. acknowledges CSIR fellowship award for this work. R.D. acknowledges funds by ANID grant FONDECYT Postdoctorado \#3220449. K.M. and S.B.P. acknowledge BRICS grant DST/IMRCD/BRICS/Pilotcall/ProFCheap/2017(G) for this work.  L.G. is grateful for financial support from the Spanish Ministerio de Ciencia e Innovaci\'on (MCIN), the Agencia Estatal de Investigaci\'on (AEI) 10.13039/501100011033, and the European Social Fund (ESF) ``Investing in your future'' under the 2019 Ram\'on y Cajal program RYC2019-027683-I and the PID2020-115253GA-I00 HOSTFLOWS project, from Centro Superior de Investigaciones Cient\'ificas (CSIC) under the PIE project 20215AT016, and the program Unidad de Excelencia Mar\'ia de Maeztu CEX2020-001058-M. L.P. acknowledges financial support from the CSIC project JAEICU-21-ICE-09. E.K. and M.S. are funded  by  a  Project 1 research grant from the Independent Research Fund Denmark (IRFD, grant number 8021-00170B) and by a VILLUM FONDEN Experiment grant (number 28021). A.V.F.'s supernova group at U.C. Berkeley is grateful for support from the TABASGO Foundation, the Christopher R.  Redlich Fund, the Miller Institute for Basic Research in Science (in which he was a Miller Senior Fellow), and many individual donors.     

\section*{Data Availability}
The photometric and spectroscopic data underlying this article will be made available upon request to the corresponding author.



\bibliographystyle{mnras}
\bibliography{sn2020jfo} 




\appendix
\section{Data tables}

\begin{table*}
\centering
\caption{{\it Swift}/UVOT photometry of SN~2020jfo$^a$}
\label{tab:phot_uvot}
\begin{tabular}{cccccccc}
\hline
Date  & Phase$^b$ & $uw2$ & $um2$ & $uw1$ & $uvu$ & $uvb$ & $uvv$ \\
(UT) & (days)& (mag) & (mag) & (mag) & (mag) & (mag) & (mag) \\
\hline
2020-05-07.13 & 3.52 &  14.75$\pm$ 0.09  &  14.75$\pm$ 0.06  &  14.64$\pm$ 0.08  &  14.55$\pm$ 0.06  &  14.74$\pm$ 0.05  &  14.99$\pm$0.06 \\
2020-05-08.73 & 5.13 &  14.81$\pm$ 0.09  &  14.66$\pm$ 0.07  &  14.60$\pm$ 0.08  &  14.40$\pm$ 0.06  &  14.53$\pm$ 0.05  &  14.62$\pm$0.06 \\
2020-05-09.23 & 5.63 &  15.18$\pm$ 0.09  &  14.86$\pm$ 0.06  &  14.79$\pm$ 0.08  &  14.42$\pm$ 0.06  &  14.58$\pm$ 0.05  &  14.62$\pm$0.06 \\
2020-05-10.35 & 6.75 &  15.39$\pm$ 0.09  &  15.08$\pm$ 0.07  &  14.87$\pm$ 0.08  &  14.40$\pm$ 0.06  &  14.48$\pm$ 0.05  &  14.54$\pm$0.06 \\
2020-05-10.42 & 6.82 &  15.47$\pm$ 0.09  &  15.12$\pm$ 0.07  &  14.97$\pm$ 0.08  &  14.48$\pm$ 0.06  &  14.53$\pm$ 0.06  &  14.58$\pm$0.07 \\
2020-05-11.22 & 7.62 &  15.76$\pm$ 0.09  &  15.34$\pm$ 0.06  &  14.96$\pm$ 0.08  &  14.41$\pm$ 0.06  &  14.49$\pm$ 0.05  &  14.51$\pm$0.06 \\
2020-05-12.76 & 9.15 &  16.31$\pm$ 0.09  &  15.91$\pm$ 0.07  &  15.36$\pm$ 0.08  &  14.54$\pm$ 0.06  &  14.59$\pm$ 0.05  &  14.58$\pm$0.06 \\
2020-05-13.42 & 9.81 &  16.33$\pm$ 0.09  &  16.05$\pm$ 0.07  &  15.48$\pm$ 0.08  &  14.57$\pm$ 0.06  &  14.52$\pm$ 0.05  &  14.58$\pm$0.06 \\
2020-05-15.67 & 12.07 &  16.83$\pm$ 0.10  &  16.73$\pm$ 0.07  &  15.95$\pm$ 0.08  &  14.78$\pm$ 0.06  &  14.58$\pm$ 0.06  &  14.75$\pm$0.07 \\
2020-05-16.66 & 13.06 &  17.17$\pm$ 0.10  &  17.05$\pm$ 0.08  &  16.18$\pm$ 0.09  &  14.84$\pm$ 0.06  &  14.66$\pm$ 0.06  &  14.73$\pm$0.07 \\
2020-05-24.91 & 21.31 &  19.93$\pm$ 0.12  &  19.90$\pm$ 0.10  &  18.20$\pm$ 0.10  &  16.23$\pm$ 0.07  &  14.96$\pm$ 0.06  &  14.66$\pm$0.06 \\
2020-05-26.56 & 22.95 &  19.78$\pm$ 0.12  &  20.20$\pm$ 0.10  &  18.53$\pm$ 0.11  &  16.62$\pm$ 0.07  &  15.04$\pm$ 0.06  &  14.71$\pm$0.06 \\
2020-06-02.76 & 30.16 &  20.89$\pm$ 0.11  &  21.34$\pm$ 0.09  &  19.30$\pm$ 0.10  &  17.33$\pm$ 0.07  &  15.45$\pm$ 0.05  &  14.79$\pm$0.06 \\
2020-06-07.11 & 34.5 &  20.60$\pm$ 0.10  &  21.24$\pm$ 0.08  &  19.51$\pm$ 0.09  &  17.71$\pm$ 0.07  &  15.57$\pm$ 0.05  &  14.86$\pm$0.05 \\
2020-06-13.09 & 40.48 &  21.19$\pm$ 0.11  &  21.79$\pm$ 0.09  &  19.96$\pm$ 0.10  &  18.12$\pm$ 0.08  &  15.72$\pm$ 0.05  &  14.93$\pm$0.06 \\
2020-06-18.00 & 45.39 &  22.29$\pm$ 0.15  &  20.90$\pm$ 0.12  &  19.66$\pm$ 0.14  &  18.65$\pm$ 0.13  &  15.90$\pm$ 0.07  &  14.91$\pm$0.07 \\
2020-07-18.41 & 75.81 &  21.50$\pm$ 0.10  &  22.63$\pm$ 0.08  &       21.83$^c$   &       20.79       &  18.58$\pm$ 0.07  &  17.51$\pm$0.08 \\
2020-07-29.33 & 86.72 &  21.74$\pm$ 0.10  &  21.74$\pm$ 0.08  &  21.82$\pm$ 0.10  &       20.52       &  18.71$\pm$ 0.07  &  17.28$\pm$0.07 \\
2020-08-03.41 & 91.81 &  21.72$\pm$ 0.10  &  21.48$\pm$ 0.08  &  21.63$\pm$ 0.10  &       20.60       &  19.00$\pm$ 0.07  &  17.53$\pm$0.08 \\
2020-08-08.45 & 96.9 &  21.73$\pm$ 0.10  &       22.14       &       21.61       &  21.49$\pm$ 0.09  &  18.86$\pm$ 0.07  &  17.47$\pm$0.10 \\
2020-11-14.65 & 195.0 &       21.86       &  22.69$\pm$ 0.10  &  21.87$\pm$ 0.10  &       20.28       &  20.28$\pm$ 0.09  &  18.58$\pm$0.15 \\
2020-12-12.03 & 222.43 &  21.64$\pm$ 0.11  &  22.05$\pm$ 0.08  &       21.67       &  21.24$\pm$ 0.09  &  21.11$\pm$ 0.08  &  19.25$\pm$0.11 \\
2020-12-24.41 & 234.81 &       22.94       &  22.14$\pm$ 0.07  &       22.17       &       21.12       &  21.41$\pm$ 0.07  &  19.55$\pm$0.08 \\
2021-01-13.53 & 254.92 &       22.98       &       22.99       &       22.24       &       21.18       &       20.37       &  20.08$\pm$0.08 \\
2021-01-28.99 & 270.39 &  22.31$\pm$ 0.14  &       21.63       &       20.99       &       20.03       &       19.29       &  20.95$\pm$0.18 \\
2021-01-30.53 & 271.92 &       22.80       &       22.83       &       22.13       &       21.10       &       20.31       &  20.47$\pm$0.09 \\

\hline
\end{tabular}
\newline\newline
$^{a}$The magnitudes of the transient have been determined after subtracting the host
flux at the SN location from recent photometry. The host flux at the SN location in
different UVOT bands has been determined from previous {\it Swift} observations (in the
year 2008) of the field (see the text for details).
\newline
$^{b}$Phase with respect to the explosion epoch (MJD = 58973.1).
\newline
$^{c}$Magnitudes without uncertainties correspond to upper limits.
\end{table*}

\begin{table*}
\centering
\caption{Details of the instruments used for the observational campaign of SN 2020jfo.}
\label{tab:instrument}
\begin{tabular}{llccccc}
\hline
Telescope   & Location  & Instrument & Pixel Scale & Imaging & Dispersers/ \\
 & & & ("/pixel) & bands & Grisms \\
\hline
0.76m KAIT & Lick Observatory & - & 0.80 & $BVRI$ & \\
1.04m ST & ARIES Observatory, India  & Tek 1k $\times$ 1k & 0.53 & $BVRI$ &  -\\
1.0m Nickel & Lick Observatory & - & 0.184 & BVRI & \\
1.0m LCO & Siding Spring Observatory & Sinistro & 0.389 & $BVg'r'i'$ & -\\
1.0m LCO & South African & Sinistro & 0.389 & $BVg'r'i'$ & -\\ 
& Astronomical Observatory & & & &\\
1.0m LCO & McDonald Observatory & Sinistro & 0.389 & $BVg'r'i'$ & -\\
1.0m LCO & Cerro Tololo & Sinistro & 0.389 & $BVg'r'i'$ & - \\ 
&  Interamerican Observatory & & & & \\
1.3m DFOT & ARIES Observatory, India & Andor $512 \times 512$ & 0.64 & $UBVRI$ & -\\
2.0m FTN & Haleakala Observatory, USA & Spectral, FLOYDS & 0.304, 0.34 & - & Cross disperser\\
2.0m FTS & Siding Spring Observatory, & FLOYDS & 0.34 & - & Cross disperser\\
& Australia & & & & \\
3.0m NASA IRTF & Mauna Kea & SpeX & 0.15 & -  & Cross disperser \\
3.0m Shane & Lick Observatory & Kast & 0.43 & - & - \\
3.6m DOT & ARIES Observatory, India & ADFOSC & - & - & - \\
10.4m GTC & La Palma, Spain & OSIRIS & 0.254 & - &  R1000B, R1000R\\
\hline
\end{tabular}
\end{table*}

\begin{figure*}
	\includegraphics[width=\textwidth]{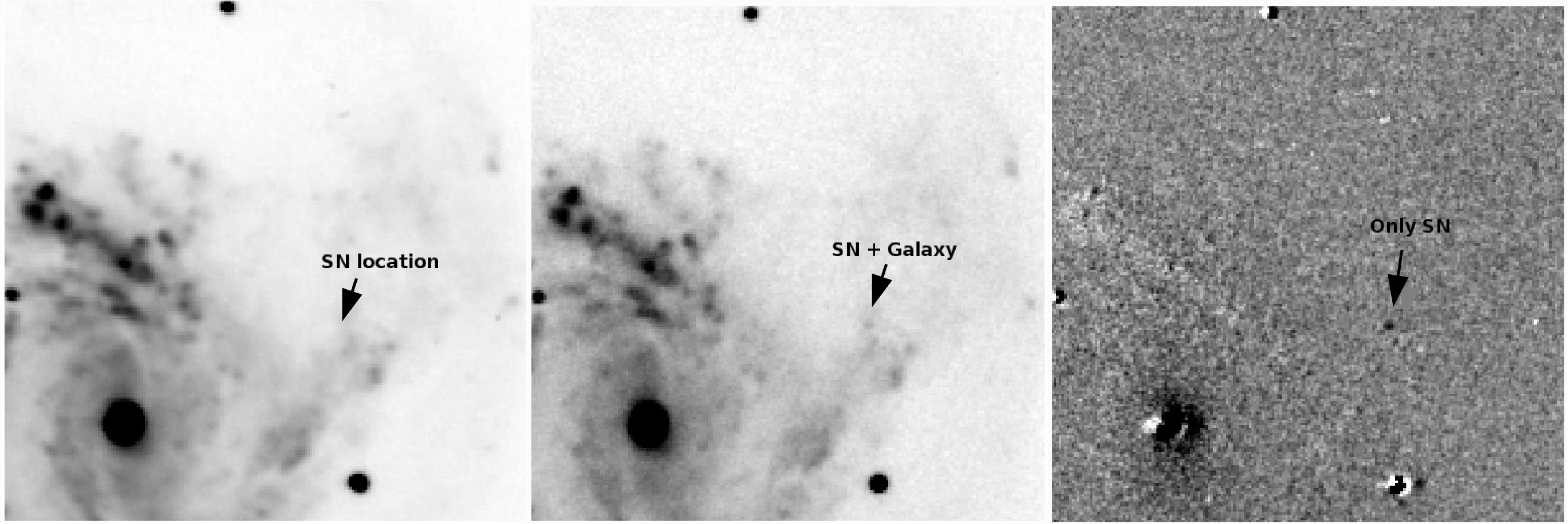}
    \caption{Measurement procedure of the true SN flux. The left and middle cutouts respectively present the template and transient frames, marking the SN locations observed from ST in the $V$ band. The template (left panel) frame was acquired on 2011-01-04 (see \citealt{roy}), whereas the middle panel represents the observation done on 2021-01-07. Both cutouts cover the same region of the sky, each of dimension $\sim2.5\arcmin \times 2.5\arcmin$, and the location of the SN has been marked. In both cutouts north is up and east is to the left. The right panel shows the subtracted frame, where the SN has been marked. We performed the template subtraction on the ST and DFOT frames using the routine outlined by \citet{roy0}. The point-spread-function (PSF) photometry was done on each of the subtracted frames. The results obtained from this routine are consistent with the photometry obtained from {\tt HOTPANTS}.}
    \label{fig:subtracted_image}
\end{figure*}

\begin{table*}
	\centering
	\caption{Optical photometry of SN 2020jfo with LCO network of telescopes}
	\begin{tabular}{ccccccccc} 
		\hline
		Date & Phase$^a$ & $B$ & $V$ & $g$ & $r$ & $i$ \\
		 (UT)   & (days)    & (mag) & (mag) & (mag) & (mag) & (mag)\\
		\hline
2020-05-06.90 & 2.80 & & & & 15.16$\pm$0.06$^b$ & \\
2020-05-07.51 & 3.41 & 14.78$\pm$0.08 & 14.77$\pm$0.09 & 14.61$\pm$0.11 & 15.02$\pm$0.05 & 15.07$\pm$0.12\\
2020-05-07.76 & 3.66 & 14.76$\pm$0.06 & 14.76$\pm$0.02 & 14.60$\pm$0.06 & 14.91$\pm$0.08 & 14.96$\pm$0.08\\
2020-05-08.44 & 4.34 & 14.67$\pm$0.02 & 14.63$\pm$0.03 & 14.66$\pm$0.50 & 14.67$\pm$0.10 & 14.86$\pm$0.01\\
2020-05-09.54 & 5.44 & 15.01$\pm$0.08 & - & 14.57$\pm$0.01 & 14.61$\pm$0.01 & 14.55$\pm$0.01\\
2020-05-10.36 & 6.26 & 14.54$\pm$0.02 & - & 14.38$\pm$0.01 & 14.42$\pm$0.04 & 14.55$\pm$0.04\\
2020-05-10.87 & 6.77 & 14.63$\pm$0.02 & 14.44$\pm$0.14 & 14.48$\pm$0.02 & 14.43$\pm$0.06 & 14.45$\pm$0.06\\
2020-05-11.56 & 7.46 & 14.61$\pm$0.05 & 14.47$\pm$0.06 & 14.23$\pm$0.02 & 14.37$\pm$0.04 & 14.51$\pm$0.04\\
2020-05-12.37 & 8.27 & 14.57$\pm$0.01 & 14.44$\pm$0.02 & 14.45$\pm$0.05 & 14.45$\pm$0.03 & 14.56$\pm$0.08\\
2020-05-12.85 & 8.75 & 14.67$\pm$0.03 & 14.53$\pm$0.02 & 14.49$\pm$0.04 & 14.56$\pm$0.06 & 14.49$\pm$0.30\\
2020-05-13.81 & 9.71 & 14.72$\pm$0.03 & 14.60$\pm$0.02 & 14.57$\pm$0.06 & 14.53$\pm$0.01 & 14.57$\pm$0.08\\
2020-05-14.83 & 10.73 & 14.82$\pm$0.04 & 14.53$\pm$0.02 & 14.55$\pm$0.01 & 14.60$\pm$0.01 & 14.69$\pm$0.02\\
2020-05-17.81 & 13.71 & 14.90$\pm$0.02 & 14.43$\pm$0.02 & 14.50$\pm$0.09 & 14.55$\pm$0.01 & 14.58$\pm$0.10\\
2020-05-19.35 & 15.25 & 14.98$\pm$0.12 & 15.29$\pm$0.04 & 13.32$\pm$0.11 & 14.81$\pm$0.02 & 14.85$\pm$0.38\\
2020-05-20.74 & 16.64 & 15.01$\pm$0.04 & 14.56$\pm$0.02 & 14.68$\pm$0.01 & 14.54$\pm$0.01 & 14.72$\pm$0.01\\
2020-05-24.84 & 20.74 & 15.25$\pm$0.02 & 14.59$\pm$0.02 & 14.80$\pm$0.01 & 14.62$\pm$0.01 & 14.67$\pm$0.01\\
2020-05-27.70 & 23.60 & 15.31$\pm$0.05 & 14.68$\pm$0.02 & 14.82$\pm$0.09 & 14.59$\pm$0.01 & 14.60$\pm$0.06\\
2020-05-29.70 & 25.60 & 15.45$\pm$0.05 & 14.63$\pm$0.02 & 14.96$\pm$0.02 & 14.58$\pm$0.01 & 14.61$\pm$0.01\\
2020-06-04.85 & 31.75 & 15.67$\pm$0.02 & 14.72$\pm$0.02 & 15.12$\pm$0.01 & 14.65$\pm$0.01 & 14.63$\pm$0.01\\
2020-06-09.75 & 36.65 & 15.83$\pm$0.02 & 14.78$\pm$0.02 & 15.23$\pm$0.01 & 14.80$\pm$0.01 & 14.74$\pm$0.01\\
2020-06-15.41 & 42.31 & 16.02$\pm$0.02 & 14.92$\pm$0.02 & 15.39$\pm$0.01 & 14.74$\pm$0.01 & 14.69$\pm$0.01\\
2020-06-19.45 & 46.35 & 15.99$\pm$0.02 & 14.80$\pm$0.02 & 15.43$\pm$0.01 & 14.68$\pm$0.01 & 14.57$\pm$0.01\\
2020-06-30.40 & 57.30 & 16.62$\pm$0.02 & 15.26$\pm$0.02 & 15.76$\pm$0.01 & - & 14.93$\pm$0.02\\
2020-07-03.80 & 60.70 & 16.75$\pm$0.02 & 15.29$\pm$0.02 & 15.93$\pm$0.01 & 14.99$\pm$0.01 & 14.91$\pm$0.01\\
2020-07-08.76 & 65.67 & 17.46$\pm$0.02 & 15.79$\pm$0.02 & 16.64$\pm$0.01 & 15.42$\pm$0.01 & 15.35$\pm$0.01\\
2020-07-15.73 & 72.63 & 18.81$\pm$0.03 & 16.99$\pm$0.02 & 17.89$\pm$0.01 & 16.49$\pm$0.01 & 16.49$\pm$0.01\\
2020-07-18.76 & 75.66 & 18.81$\pm$0.03 & - & - & - & -\\
2020-07-20.71 & 77.61 & 19.00$\pm$0.02 & 17.28$\pm$0.02 & 17.97$\pm$0.01 & 16.61$\pm$0.01 & 16.58$\pm$0.01\\
2020-07-23.74 & 80.64 & 18.68$\pm$0.02 & 17.11$\pm$0.03 & 17.77$\pm$0.02 & 16.51$\pm$0.01 & 16.49$\pm$0.02\\
2020-07-27.76 & 84.66 & 18.84$\pm$0.14 & 17.37$\pm$0.06 & 17.72$\pm$0.01 & 16.47$\pm$0.02 & 16.71$\pm$0.03\\
2020-07-31.38 & 88.28 & - & 17.78$\pm$0.39 & - & 17.17$\pm$0.18 & 17.13$\pm$0.35\\
2020-08-05.74 & 93.64 & 18.70$\pm$0.06 & 17.30$\pm$0.04 & 18.03$\pm$0.03 & 16.58$\pm$0.02 & 16.90$\pm$0.03\\
2020-08-10.37 & 98.27 & 18.51$\pm$0.09 & 17.41$\pm$0.09 & 17.83$\pm$0.03 & 16.59$\pm$0.04 & 16.91$\pm$0.03\\
2020-12-19.48 & 229.38 & 20.04$\pm$0.06 & 19.03$\pm$0.04 & 19.35$\pm$0.03 & 18.45$\pm$0.02 & 18.53$\pm$0.04\\
2020-12-23.45 & 233.35 & 20.16$\pm$0.07 & 19.12$\pm$0.08 & 19.44$\pm$0.03 & 18.61$\pm$0.02 & 18.91$\pm$0.05\\
2021-01-03.50 & 244.40 & 20.57$\pm$0.10 & 19.57$\pm$0.10 & 19.85$\pm$0.08 & 18.97$\pm$0.06 & 19.09$\pm$0.13\\
2021-01-15.07 & 255.97 & 20.44$\pm$0.08 & 19.41$\pm$0.06 & 19.54$\pm$0.03 & 19.12$\pm$0.04 & 18.85$\pm$0.06\\
2021-01-26.08 & 266.98 & 20.57$\pm$0.08 & 19.54$\pm$0.07 & 19.84$\pm$0.03 & 19.62$\pm$0.04 & 19.46$\pm$0.03\\
2021-02-06.08 & 277.98 & 20.90$\pm$0.11 & 19.68$\pm$0.06 & - & 19.66$\pm$0.05 & -\\
2021-02-09.34 & 281.24 & 21.20$\pm$0.16 & 19.83$\pm$0.06 & 19.92$\pm$0.04 & 19.72$\pm$0.06 & 19.39$\pm$0.07\\
2021-02-27.66 & 299.56 & 21.08$\pm$0.14 & 20.28$\pm$0.08 & 19.92$\pm$0.05 & 20.33$\pm$0.08 & 19.80$\pm$0.07\\
2021-03-11.23 & 311.13 & 21.69$\pm$0.28 & 20.32$\pm$0.09 & 20.29$\pm$0.05 & 20.46$\pm$0.13 & 19.85$\pm$0.08\\
2021-03-22.24 & 322.14 & 21.63$\pm$0.30 & 20.51$\pm$0.09 & 20.46$\pm$0.06 & 20.79$\pm$0.15 & 20.09$\pm$0.07\\
2021-04-03.95 & 334.85 & 22.24$\pm$0.34 & 20.66$\pm$0.14 & 20.58$\pm$0.06 & 21.44$\pm$0.18 & 20.56$\pm$0.13\\
2021-04-14.89 & 345.79 & 22.50$\pm$0.45 & 20.66$\pm$0.13 & 20.61$\pm$0.07 & 21.49$\pm$0.19 & 20.47$\pm$0.19\\
2021-04-27.18 & 358.08 & 21.53$\pm$0.13 & 21.60$\pm$0.39 & - & - & 20.88$\pm$0.38\\
2021-05-08.18 & 369.08 & 22.61$\pm$0.38 & 21.07$\pm$0.18 & 21.43$\pm$0.39 & - & -\\
2021-05-13.36 & 374.26 & - & 21.03$\pm$0.15 & - & - & -\\
2021-05-24.50 & 385.40 & 22.00$\pm$0.38 & 21.51$\pm$0.32 & 22.05$\pm$0.22 & 21.38$\pm$0.40 & 20.28$\pm$0.22\\
2021-06-04.71 & 396.61 & - & 21.18$\pm$0.21 & 21.64$\pm$0.16 & - & -\\
2021-06-06.41 & 398.31 & - & 21.34$\pm$0.24 & 21.30$\pm$0.12 & 21.92$\pm$0.28 & 21.48$\pm$0.44\\
2021-06-21.06 & 412.96 & 22.68$\pm$0.51 & 22.50$\pm$0.43 & - & 21.83$\pm$0.23 & 21.15$\pm$0.23\\
2021-07-02.04 & 423.94 & - & 21.53$\pm$0.31 & 21.91$\pm$0.26 & 22.01$\pm$0.34 & -\\
2021-07-13.00 & 434.90 & - & - & 21.52$\pm$0.50 & - & -\\
		\hline
	\end{tabular}
	\newline\newline
$^{a}$Phase with respect to the explosion epoch (MJD = 58973.1).\\
$^{b}$The earliest photometric data obtained with the 10.4 m GTC.
	\label{tab:lco_phot}
\end{table*}
\begin{table*}
	\centering
    \caption{Optical photometry of SN 2020jfo with ARIES telescopes}
	\begin{tabular}{ccccccccc} 
		\hline
		Date & Phase$^a$ & $B$ & $V$ & $R$ & $I$ \\
		 (UT)   &  (days)   & (mag) & (mag) & (mag) & (mag) \\
		\hline
2021-01-06.98  & 248.37 &       $-$      & 19.26$\pm$0.05 & 19.09$\pm$0.09 & 19.89$\pm$0.17 \\
2021-01-07.90  & 249.30 & 19.78$\pm$0.09 & 19.20$\pm$0.06 & 19.31$\pm$0.09 & 19.55$\pm$0.20 \\
2021-01-07.91  & 249.31 &       $-$      & 19.19$\pm$0.05 & 19.50$\pm$0.06 & 19.49$\pm$0.16 \\
2021-01-09.90  & 251.26 & 20.14$\pm$0.21 & 19.21$\pm$0.05 & 19.20$\pm$0.08 & 19.66$\pm$0.16 \\
2021-01-21.96  & 263.34 & 20.07$\pm$0.19 & 19.11$\pm$0.06 & 18.84$\pm$0.08 & 19.97$\pm$0.28 \\
2021-02-02.98  & 275.38 &       20.4     & 19.77$\pm$0.14 & 18.81$\pm$0.10 & 19.86$\pm$0.27 \\
2021-02-14.79  & 287.19 & 20.59$\pm$0.23 & 20.16$\pm$0.09 & 18.90$\pm$0.08 & 19.98$\pm$0.17 \\
2021-02-22.81  & 295.37 & 21.21$\pm$0.32 & 20.17$\pm$0.10 & 19.67$\pm$0.13 & 21.12$\pm$0.32 \\
		\hline
	\end{tabular}
\newline\newline
$^{a}$Phase with respect to the explosion epoch (MJD = 58973.1).
\newline
$^{b}$Magnitude without error corresponds to the upper limit.
	\label{tab:aries_phot}
\end{table*}
\begin{table*}
	\centering
	
	\caption{Optical photometry of SN 2020jfo with Lick Observatory telescopes}
	\begin{tabular}{cccccccccccc} 
		\hline
		Date & Phase$^a$ & $B$ & $V$ & $R$ & $I$ & {\it Clear}\\
		 (UT)   & (days)    & (mag) & (mag) & (mag) & (mag) & (mag) \\
		\hline
2020-05-07.32 &	3.22 &	14.66$\pm$0.06 &	15.01$\pm$0.03 &	14.94$\pm$0.06 &	14.96$\pm$0.06 &	14.78$\pm$0.04\\
2020-05-08.35 &		4.25 &		14.58$\pm$0.04 &		14.77$\pm$0.02 &		14.71$\pm$0.02 &		14.64$\pm$0.03 &		14.55$\pm$0.03\\
2020-05-09.25 &		5.15 &		14.52$\pm$0.04 &		14.66$\pm$0.02 &		14.58$\pm$0.02 &		14.47$\pm$0.02 &		14.44$\pm$0.03\\
2020-05-10.34 &		6.24 &		14.54$\pm$0.05 &		14.63$\pm$0.02 &		14.48$\pm$0.02 &		14.37$\pm$0.03 &		14.38$\pm$0.02\\
2020-05-11.30 &		7.20 &		14.48$\pm$0.05 &		14.59$\pm$0.03 &		14.45$\pm$0.02 &		14.31$\pm$0.03 &		14.37$\pm$0.02\\
2020-05-16.27 &		12.17 &		14.66$\pm$0.03 &		14.71$\pm$0.01 &		14.51$\pm$0.01 &		14.45$\pm$0.02 &		14.44$\pm$0.02\\
2020-05-17.27 &		13.17 &		14.68$\pm$0.05 &		14.76$\pm$0.04 &		14.50$\pm$0.04 &		14.42$\pm$0.10 &		14.46$\pm$0.02\\
2020-05-21.24 &		17.14 &		14.79$\pm$0.08 &		14.66$\pm$0.02 &		14.48$\pm$0.02 &		14.37$\pm$0.03 &		 -\\
2020-05-21.31 &		17.21	 &	14.87$\pm$0.04 &		14.73$\pm$0.02 &		14.51$\pm$0.02 &		14.43$\pm$0.03 &		14.46$\pm$0.03\\
2020-05-23.25 &		19.15 &		14.94$\pm$0.04 &		14.71$\pm$0.02 &		14.50$\pm$0.02 &		14.38$\pm$0.02 &		14.48$\pm$0.03\\
2020-05-24.27 &		20.17 &		15.01$\pm$0.04 &		14.73$\pm$0.02 &		14.50$\pm$0.02 &		14.39$\pm$0.03 &		14.49$\pm$0.03\\	
2020-05-25.28 &		21.18 &		15.00$\pm$0.08 &		14.77$\pm$0.02 &		14.52$\pm$0.02 &		14.41$\pm$0.02 &		14.49$\pm$0.03\\
2020-05-26.25 &		22.15 &		15.10$\pm$0.04 &		14.75$\pm$0.02 &		14.51$\pm$0.02 &		14.38$\pm$0.03 &		14.51$\pm$0.02\\
2020-05-27.25 &		23.15 &		15.14$\pm$0.04 &		14.77$\pm$0.02 &		14.53$\pm$0.02 &		14.41$\pm$0.03 &		14.52$\pm$0.03\\
2020-05-28.26 &		24.16 &		15.20$\pm$0.04 &		14.78$\pm$0.02 &		14.52$\pm$0.02 &		14.37$\pm$0.02 &		14.55$\pm$0.03\\
2020-06-02.18 &		29.08 &		15.39$\pm$0.09 &		14.78$\pm$0.08 &		14.54$\pm$0.13 &		14.26$\pm$0.10	 &	 -\\
2020-06-02.24 &		29.14 &		15.35$\pm$0.09 &		14.76$\pm$0.03 &		14.54$\pm$0.02 &		14.34$\pm$0.03 &		 -\\
2020-06-04.27 &		31.17 &		15.50$\pm$0.05 &		14.83$\pm$0.02 &		14.56$\pm$0.02 &		14.38$\pm$0.03 &		14.60$\pm$0.02\\
2020-06-06.26 &		33.16 &		15.47$\pm$0.13 &		14.79$\pm$0.06 &		14.54$\pm$0.02 &		14.32$\pm$0.04 &		 -\\
2020-06-10.21 &		37.11 &		15.67$\pm$0.08 &		14.86$\pm$0.03 &		14.59$\pm$0.02 &		14.35$\pm$0.03 &		 -\\
2020-06-14.24 &		41.14 &		15.74$\pm$0.08 &		14.88$\pm$0.03 &		14.60$\pm$0.02 &		14.36$\pm$0.03 &		 -\\
2020-06-18.24 &		45.14 &		15.91$\pm$0.23 &		14.92$\pm$0.14 &		14.65$\pm$0.02 &		14.38$\pm$0.04 &		 -\\
2020-06-22.25 &		49.15 &		16.05$\pm$0.08	 &	14.97$\pm$0.03 &		14.66$\pm$0.02	 &	14.42$\pm$0.25 &		 -\\
2020-06-30.20 &		57.10 &		16.27$\pm$0.08 &		15.20$\pm$0.03 &		14.82$\pm$0.02	 &	14.55$\pm$0.04	 &	 -\\
2020-06-30.24 &		57.14 &		16.27$\pm$0.07 &		15.23$\pm$0.03 &		14.84$\pm$0.03 &		14.57$\pm$0.03 &		14.88$\pm$0.04\\
2020-07-01.21 &		58.11 &		16.34$\pm$0.07 &		15.29$\pm$0.03 &		14.91$\pm$0.03 &		14.63$\pm$0.04 &		14.99$\pm$0.04\\
2020-07-04.23 &		61.13 &		16.58$\pm$0.08 &		15.43$\pm$0.02	 &	14.97$\pm$0.02 &		14.71$\pm$0.03 &		15.08$\pm$0.04\\
2020-07-05.20 &		62.10 &		16.78$\pm$0.10 &		15.57$\pm$0.03 &		15.08$\pm$0.03 &		14.78$\pm$0.03 &		15.20$\pm$0.04\\
2020-07-08.22 &		65.12 &		17.16$\pm$0.09 &		15.83$\pm$0.04 &		15.30$\pm$0.03 &		15.00$\pm$0.04 &		15.45$\pm$0.05\\
2020-07-10.20 &		67.10 &		17.44$\pm$0.13 &		16.18$\pm$0.05 &		15.57$\pm$0.05 &		15.30$\pm$0.05 &		15.76$\pm$0.06\\
2020-07-11.22 &		68.12 &		17.68$\pm$0.15 &		16.43$\pm$0.05 &		15.76$\pm$0.04 &		15.45$\pm$0.05 &		15.91$\pm$0.07\\
2020-07-18.20 &		75.10 &		 - & - &		 - &		 - &		16.55$\pm$0.20\\
2020-07-19.20 &		76.10 &		17.96$\pm$0.24 &		17.20$\pm$0.10 &		16.43$\pm$0.08 &		16.06$\pm$0.07 &		16.53$\pm$0.12\\
2020-07-20.20 &		77.10 &		 - &		 - &		 - &		 - &		16.49$\pm$0.13\\
2020-07-21.20 &		78.10 &		18.20$\pm$0.28 &		17.14$\pm$0.10 &		16.41$\pm$0.07 &		16.10$\pm$0.07 &		16.52$\pm$0.11\\
2020-07-23.20 &		80.10 &		18.26$\pm$0.38 &		17.18$\pm$0.15 &		16.37$\pm$0.11 &		16.15$\pm$0.09 &		16.55$\pm$0.14\\

		\hline
	\end{tabular}
	\newline\newline
$^{a}$ Phase with respect to the explosion epoch (MJD = 58973.1).
	\label{tab:shane_phot}
\end{table*}
\begin{table}
	\centering
	\caption{Spectroscopic log of SN~2020jfo}
	
	\begin{tabular}{lcl} 
		\hline
		Date & Phase$^a$ &Telescope\\
		  (UT) & (days) &  \\
		\hline
	&	Optical &\\
	\hline
2020-05-06.90  & 2.80 &  GTC \\
2020-05-07.54  & 3.44 &  FTS \\
2020-05-08.55  & 4.45 &  FTS \\
2020-05-12.36  & 8.26 &  FTN \\
2020-05-16.49  & 12.39 &  FTS \\
2020-05-17.27  & 13.17 &  Shane \\
2020-05-19.50  & 15.40 &  FTS \\
2020-05-23.00  & 18.90 &  DOT \\
2020-05-24.00  & 19.90 &  DOT \\
2020-05-24.46  & 20.36 &  FTS \\
2020-05-25.33  & 21.17 &  Shane\\
2020-05-28.48  & 24.38 &  FTS \\
2020-05-29.29  & 25.19 &  Shane \\
2020-06-05.48  & 32.38 &  FTS \\
2020-06-13.30  & 40.20 &  FTN \\
2020-06-26.27   & 53.17 & FTN \\
2020-07-06.42  & 63.32 &  FTS \\
2020-07-09.20  & 66.10 &  Shane\\
2020-07-14.26  & 71.16 &  FTN \\
2020-07-18.20  & 75.09 &  Shane \\
2020-07-23.25  & 80.15 &  FTN \\
2020-07-28.19  & 85.09 &  Shane \\
2020-08-01.35   &  89.25 & FTS \\
2021-01-13.00  & 253.90 &  DOT \\
2021-01-15.00  & 255.90 &  DOT\\
2021-05-04.30  & 365.20 &  Shane \\
\hline
 & NIR & \\
 \hline
2020-05-13 & 9.27 & IRTF\\
2020-05-24 & 20.24 & IRTF\\
2020-06-15 & 42.19 & IRTF\\ 
		\hline
	\end{tabular}
		\newline\newline
$^{a}$ Phase with respect to the explosion epoch (MJD = 58973.1).
	\label{tab:spec_log}
\end{table}


\bsp	
\label{lastpage}
\end{document}